\documentclass[12pt]{article}

\usepackage{latexsym,amsmath,amscd,amssymb,graphics,diagrams,setspace}
\usepackage{enumerate,framed}

\usepackage{graphicx}
\usepackage{epstopdf}

\usepackage[square,authoryear]{natbib}
\usepackage[colorlinks]{hyperref}
\usepackage{url}

\usepackage[all]{xy}

\makeatletter

\@addtoreset{figure}{section}
\def\thefigure{\thesection.\@arabic\c@figure}
\def\fps@figure{h, t}
\@addtoreset{table}{bsection}
\def\thetable{\thesection.\@arabic\c@table}
\def\fps@table{h, t}
\@addtoreset{equation}{section}

\makeatother

\newcommand{\beq}{\begin{eqnarray*}}
\newcommand{\eeq}{\end{eqnarray*}}

\newcommand{\rem}[1]{}

\newcommand{\mg}{{\mathfrak{g}}}

\newcommand{\cA}{{\mathcal{A}}}
\newcommand{\cB}{{\mathcal{B}}}
\newcommand{\cC}{{\mathcal{C}}}

\newcommand{\cF}{{\mathcal{F}}}

\newcommand{\cH}{{\mathcal{H}}}

\newcommand{\cK}{{\mathcal{K}}}
\newcommand{\cL}{{\mathcal{L}}}

\newcommand{\cN}{{\mathcal{N}}}

\newcommand{\cR}{{\mathcal{R}}}

\newcommand{\mI}{{\mathbb{I}}}

\newcommand{\mR}{{\mathbb{R}}}

\newcommand{\al}{\alpha}
\newcommand{\be}{\beta}
\newcommand{\gam}{\gamma}

\newcommand{\de}{\delta}

\newcommand{\sig}{\sigma}
\newcommand{\Sig}{\Sigma}

\newcommand{\om}{\omega}
\newcommand{\Om}{\Omega}

\newcommand{\br}{{\boldsymbol{r}}}

\newcommand{\brho}{{\boldsymbol{\rho}}}

\newcommand{\mso}{{\mathfrak{so}}}

\newcommand{\msu}{{\mathfrak{su}}}

\newcommand{\half}{{\frac{1}{2}}}

\newcommand{\pp}[2]{\frac{\partial #1}{\partial #2}}

\newcommand{\dd}[2]{\frac{d #1}{d #2}}

\newcommand{\prt}{\partial}
\newcommand{\dede}[2]{\frac{\de #1}{\de #2}}
\newcommand{\DD}[2]{\frac{D #1}{D #2}}

\newcommand{\eval}[2]{{\left. #1 \right|_{#2}}}

\newcommand{\Ad}{{\operatorname{Ad}}}
\newcommand{\ad}{{\operatorname{ad}}}
\newcommand{\tr}{{\operatorname{tr}}}

\newcommand{\id}{{\operatorname{id}}}

\newcommand{\hor}{{\operatorname{Hor}}}

\newcommand{\dive}{{\operatorname{div}}}

\def\contract{\makebox[1.2em][c]{\mbox{\rule{.6em}
{.01truein}\rule{.01truein}{.6em}}}}

\newcommand{\lsb}{\left[}
\newcommand{\llsb}{ \left[ \! \left[}
\newcommand{\rsb}{\right]}
\newcommand{\rrsb}{\right ] \! \right ]}

\newcommand{\lp}{\left(}
\newcommand{\rp}{\right)}

\newcommand{\scp}[2]{{\left\langle {#1}\, , \, {#2}\right\rangle}}
\newcommand{\dscp}[2]{{\left\langle \!\left\langle {#1}\, , \, {#2}\right\rangle \!\right\rangle}}

\newtheorem{theorem}{Theorem}[section]

\newtheorem{lemma}[theorem]{Lemma}
\newtheorem{remark}[theorem]{Remark}
\newtheorem{proposition}[theorem]{Proposition}
\newtheorem{corollary}[theorem]{Corollary}

\newenvironment{proof}[1][Proof]{\textbf{#1.} }{\ \rule{0.5em}{0.5em}}

\begin{document}



\title{Lagrange-Poincar\'e field equations}
\author{David C.P. Ellis${}^1$, Fran\c{c}ois Gay-Balmaz$^{2}$, Darryl D. Holm$^{1}$,  and Tudor S. Ratiu$^{3}$}
\addtocounter{footnote}{1}
\footnotetext{Department of Mathematics, Imperial College London. London SW7 2AZ, UK. Partially supported by Royal Society of London, Wolfson Award.
\texttt{david.ellis102@imperial.ac.uk, d.holm@imperial.ac.uk}
\addtocounter{footnote}{1} }
\footnotetext{Control and Dynamical Systems, California Institute of Technology 107-81, Pasadena, CA 91125, USA and Laboratoire de 
M\'et\'eorologie Dynamique, \'Ecole Normale Sup\'erieure/CNRS, Paris, France. Partially supported by a Swiss NSF grant.
\texttt{fgbalmaz@cds.caltech.edu}
\addtocounter{footnote}{1} }
\footnotetext{Section de
Math\'ematiques and Bernoulli Center, \'Ecole Polytechnique F\'ed\'erale de
Lausanne.
CH--1015 Lausanne. Switzerland. Partially supported by a Swiss NSF grant.
\texttt{Tudor.Ratiu@epfl.ch}
\addtocounter{footnote}{1} }

\date{}
\maketitle

\makeatother

\maketitle


\noindent \textbf{AMS Classification:} 70S05, 70S10

\noindent \textbf{Keywords:} field theories, symmetries, covariant reduction, Euler-Lagrange equations, conservation laws

\begin{abstract}

The Lagrange-Poincar\'e equations of classical mechanics are cast into a field theoretic context together with their associated constrained variational principle.  An integrability/reconstruction condition is established that relates solutions of the original problem with those of the reduced problem.  The Kelvin-Noether theorem is formulated in this context.  Applications to the isoperimetric problem, the Skyrme model for meson interaction, metamorphosis image dynamics, and molecular strands illustrate various aspects of the theory.

\end{abstract}

\tableofcontents


\section{Introduction}\label{sec:Intro}

Reduction by symmetry of Lagrangian field theories has aided the implementation of many diverse mathematical models from geometric mechanics.  Two main approaches have been developed.  One approach, investigated in \cite{GIMMSYI}, employs multisymplectic geometry to extend the symplectic formulation of classical Lagrangian systems.  The second approach, studied in \cite{CaRaSh2000} and \cite{CaRa2003}, reduces the variational principal itself without reference to the Hamiltonian side and is referred to as \textit{covariant Lagrangian reduction}.

A comparison of the covariant Lagrangian reduction approach, \cite{CaRaSh2000, CaRa2003}, with the corresponding classical Lagrangian reduction method, \cite{HoMaRa1998, CeHoMaRa1998, CeMaRa2001}, shows that a paradigm permitting both reductions is currently lacking. Further, examples such as the isoperimetric problem and metamorphosis image dynamics require such a paradigm.  And further still, the desired capability for covariant reformulations of classical problems, such as in \cite{MaSh1999}, call for such a paradigm.

This paper achieves a full generalization of the classical theory while preserving the flavor of the current covariant theory.  Applications to the \textit{Skyrme model} and the \textit{molecular strand} illustrate the ideas of \cite{CaRaSh2000} and \cite{CaRa2003} in \S\ref{sec:Principal} and \S\ref{sec:Subgroup}, respectively.  A general discussion of classical Lagrangian reduction appears in \S\ref{sec:fibre}.  These discussions illustrate the need for the development of the more general theory.

\subsection{Principal bundle reduction: the Skyrme model}\label{sec:Principal}
The first in the series of papers on covariant Lagrangian reduction, \cite{CaRaSh2000}, dealt with the extension of classical Euler-Poincar\'e reduction of variational principles to the field theoretic context.  There, a field theory was formulated on a principal bundle and was reduced by the structure group.  These results may be illustrated by the \textit{Skyrme model} for pion interaction, which was first developed in \cite{Sk1961} and whose more recent developments were reviewed from a Physics-based standpoint in \cite{GiPa1998}.

Both the original formulation of the classical Skyrme model and its recent advances have been described, for example in \cite{GiPa1998}, in terms of local coordinates. The use of local coordinates, while necessary for numerical implementation, tends to obscure the geometric content of the equations.  Therefore this paper approaches the theory, on the whole, from a coordinate-free viewpoint.  However, to aid communication and compatibility with the references, some of the examples are addressed in local coordinates too.

\paragraph{The Skyrme model.}

The class of model given in \cite{GiPa1998} may be outlined as follows:

Consider a unitary field $u:S^3 \to SU(n)$ over the three-sphere $S^3$ for $n$ either $2$ or $3$.  The three-sphere $S^3$ is interpreted as a one-point compactification of Euclidean space $\mR^3$.  In local coordinates, the massless Skyrme Lagrangian reads
\begin{equation}\label{L_Sky}
L_{Sky}\lp u, \prt_j u\rp = -\frac{f_\pi}{4}\tr \lp u^\dagger (\prt_j u) u^\dagger (\prt_j u)\rp + \frac{1}{32e^2}\tr\lp \lsb u^\dagger (\prt_ju), u^\dagger (\prt_k u)\rsb^2\rp
\end{equation}
where $u^\dagger := \bar u^T$ is the adjoint of $u$.  The constants $f_\pi$ and $e^2$ are potentially calculable from QCD but, in practice, are fitted to experimental data.  The local representation of the Euler-Lagrange equations for $L_{Sky}$ are given by
\begin{equation}
\prt_j\lp \frac{f_\pi}{2}(\prt^ju) u^\dagger + \frac{1}{8e^2}\lsb (\prt_ku) u^\dagger, \lsb (\prt^ju)u^\dagger, (\prt^ku) u^\dagger\rsb \rsb \rp = 0.
\label{eq:ELSky}
\end{equation}
Baryons are identified with topological soliton solutions of equation \eqref{eq:ELSky} with $u:S^3 \to SU(n)$.  Note that the Lagrangian $L_{Sky}$ is $SU(n)$-invariant under the transformation
\[
(u,\partial_ju)\mapsto \left(vu,v(\partial_ju) \right),\quad\text{for all}\quad v\in SU(n).
\]
Therefore, a reduction by symmetry may be effected.  In order to bring out the geometry of the system a reformulation of the problem is required.

\paragraph{Geometric formulation.} Let $\pi:=\pi_{S^3,P}:P\to S^3$ be a principal $SU(n)$ bundle over the three-sphere $S^3$.  A \textit{section} of $\pi$ is a smooth map $\sig: S^3 \to P$, such that 
\begin{equation}
\pi\circ \sig = \id_{S^3},
\label{eq:DefSec}
\end{equation}
where $\id_{S^3}$ is the identity map on $S^3$. The space of sections of $\pi$ is denoted $\Gamma\lp\pi\rp$. Recall that a principal bundle admits a section if and only if it is trivial. Therefore, in general only \textit{local sections} $\sigma:U\subset S^3\rightarrow P$ defined on an open subset $U\subset S^3$ may be considered.

In a local trivialization $U\subset S^3$, a section  $\sigma$ reads $\sig(x) = (x, u(x))$, where $u:U\subset S^3\rightarrow SU(n)$. Thus, the space of sections $\Gamma\lp \pi\rp$ corresponds to the space of unitary fields.

Recognize that $\lp u, \prt_ju\rp$ is a local representation of the tangent map $T\sig:TS^3\rightarrow TP$.  The \textit{jet bundle}, $J^1P$, provides the natural space to consider such objects. This affine bundle over $P$ may be defined fiberwise by
\[
J^1_pP = \left\{ \eval{\gamma_p \in L\lp T_xS^3, T_pP\rp}{} T_p\pi\circ \gamma_p = \id_{T_xS^3}\right\},
\]
with projection $\pi_{P,J^1P}: J^1P \to P$ given by $\pi_{P,J^1P}\lp\gamma_p\rp = p$.  The jet bundle serves field theories as the tangent bundle serves classical Lagrangian systems.

For the most part, $J^1P$ is considered as a fiber bundle over $S^3$ with projection
\[
\pi_{S^3,J^1P}:=\pi\circ \pi_{P,J^1P}:J^1P \to S^3.
\]
Indeed, the tangent map of a section $\sigma\in\Gamma(\pi)$, interpreted as a map $x\mapsto T_x\sigma$, offers a section of $\pi_{S^3,J^1P}$ since for $T_x\sig \in L\lp T_xS^3, T_{\sig(x)}P\rp$ equation \eqref{eq:DefSec} yields
\[
T_{\sig(x)}\pi\circ T_x\sig = T_x \lp \pi\circ \sig\rp = T_x \id_{S^3} = \id_{T_xS^3}.
\]

The geometry introduced here is succinctly visualized and organized by commutative diagrams.  The following commutative diagram exhibits the geometry of the jet bundle:
\begin{diagram}
J^1P &                      & \rTo^{\boldsymbol{\pi_{S^3,J^1P}}} &             & S^3 \\
     & \rdTo_{\boldsymbol{\pi_{P,J^1P}}} &                     & \ruTo_{\boldsymbol{\pi}} &   \\
     &                      & P                   &             &   \\
\end{diagram}
Arrows between spaces indicate maps from the space at the tail of the arrow to the space at its head. Sometimes arrows are adorned with the name of the maps they represent. Different paths through the diagram are equivalent in terms of composition of the associated maps; therefore, this diagram also communicates the relation
\[
\pi_{S^3, J^1P} = \pi\circ \pi_{P, J^1P}.
\]

\paragraph{The reduced bundle.} Having identified the geometry of the Classical Skyrme Model, one may proceed by thinking about reduction by left $SU(n)$ symmetry in the style of \cite{CaRaSh2000}.
The quantities $(\prt_ju)u^\dagger$ may be understood as the local representation of a principal connection form $\cA\in\Omega^1(P,\msu(n))$ which has been pulled back by the unitary field where $\msu(n)$ denotes the Lie algebra of $SU(n)$.
The connection form $\cA\in\Omega^1(P,\mathfrak{u}(n))$ on $P$ provides the required geometric tool to effect the reduction since it provides a vector bundle isomorphism
\begin{equation}\label{isom_skyrme}
J^1P/SU(n) \to L\lp TS^3, \Ad P\rp,
\end{equation}
where $\Ad P$ denotes the \textit{adjoint bundle} associated to the principal bundle $P$ defined as the quotient space
\[
\Ad P := \lp P \times \msu(n)\rp/SU(n),
\]
relative to the following diagonal action of $u\in SU(n)$:
\[
(p,\xi)\in P\times \msu(n) \mapsto \left(up,\Ad_{u}\xi\right)\in P\times \msu(n).
\]
Denoting the equivalence class of $(p,\xi) \in P\times \msu(n)$ by
\[
\llsb p, \xi \rrsb_{\mathfrak{su}(n)} \in \Ad P,
\]
the bundle isomorphism $J^1P/SU(n) \rightarrow L\lp TS^3, \Ad P\rp$ reads
\[
\lsb T\sig \rsb \mapsto \bar\sig:=\llsb \sig, \sig^*\cA\rrsb_{\mathfrak{su}(n)}.
\]

\paragraph{The reduced Euler-Lagrange equations.} The Skyrme model is now written in the same form as the result of \cite{CaRaSh2000} which states that the Euler-Lagrange equations on $J^1P$ are equivalent to the covariant Euler-Poincar\'e equations on $L\lp TS^3,\Ad P\rp$, which read
\begin{equation}\label{cov_Euler_P}
\dive^\cA\dede{l}{\bar\sig} - \ad^*_{\bar\sig}\dede{l}{\bar\sig} =0.
\end{equation}
Here $\dive^\cA$ denotes the divergence associated to the covariant exterior derivative on $\Ad P$ associated with the principal connection $\cA$ and $\ad^*$ is the dual of the adjoint operator on $\msu(n)$.  For the classical Skyrme model, the reduced Lagrangian associated to \eqref{L_Sky} can be written as
\[
l(\bar\sigma)=\frac{1}{2}\|\bar\sigma\|^2+\frac{1}{4}\|\bar\sigma\wedge\bar\sigma\|^4,
\]
where $\|\cdot \|$ is the norm associated with a Riemannian metric on $S^3$ and an Ad-invariant inner product on $\mathfrak{g}$.
The unreduced Lagrangian density is
\[
\mathcal{L}(\gamma_p)=\frac{1}{2}\|\llsb p,\mathcal{A}\circ\gamma_p\rrsb_{\msu(n)}\|^2+\frac{1}{4}\|\llsb p,\mathcal{A}\circ\gamma_p\rrsb_{\msu(n)}\wedge\llsb p,\mathcal{A}\circ\gamma_p\rrsb_{\msu(n)} \|^2
\]
and is clearly $SU(n)$-invariant.
The classical Skyrme model equations then become
\begin{equation}
\dive^\cA\bar\Pi - \ad^*_{\bar\sig}\bar\Pi=0 \qquad \bar\Pi = \frac{f_\pi}{2}\bar\sig^\flat + \frac{1}{8e^2} \ad^*_{\bar\sig} \lp\bar\sig \wedge \bar\sig\rp^\flat.
\label{eq:LPSky}
\end{equation}
where $\flat$ denotes the flat map $\flat:L(TS^3,\Ad P) \to L(T^*S^3,\Ad^* P)$ induced by the Riemannian metric on $S^3$ and the Ad-invariant inner product on $\mathfrak{g}$.
The local representation of equation \eqref{eq:LPSky} is equation \eqref{eq:ELSky}.  For details on related dynamical systems to the classical Skyrme model see \cite{Ho2008b}. The link between the covariant and dynamical reductions associated to the equation \eqref{cov_Euler_P} is established in \cite{GBRa2009}.

\subsection{Subgroup reduction: the molecular strand}\label{sec:Subgroup}
\begin{figure} [h]

\centering

\includegraphics[width=12cm]{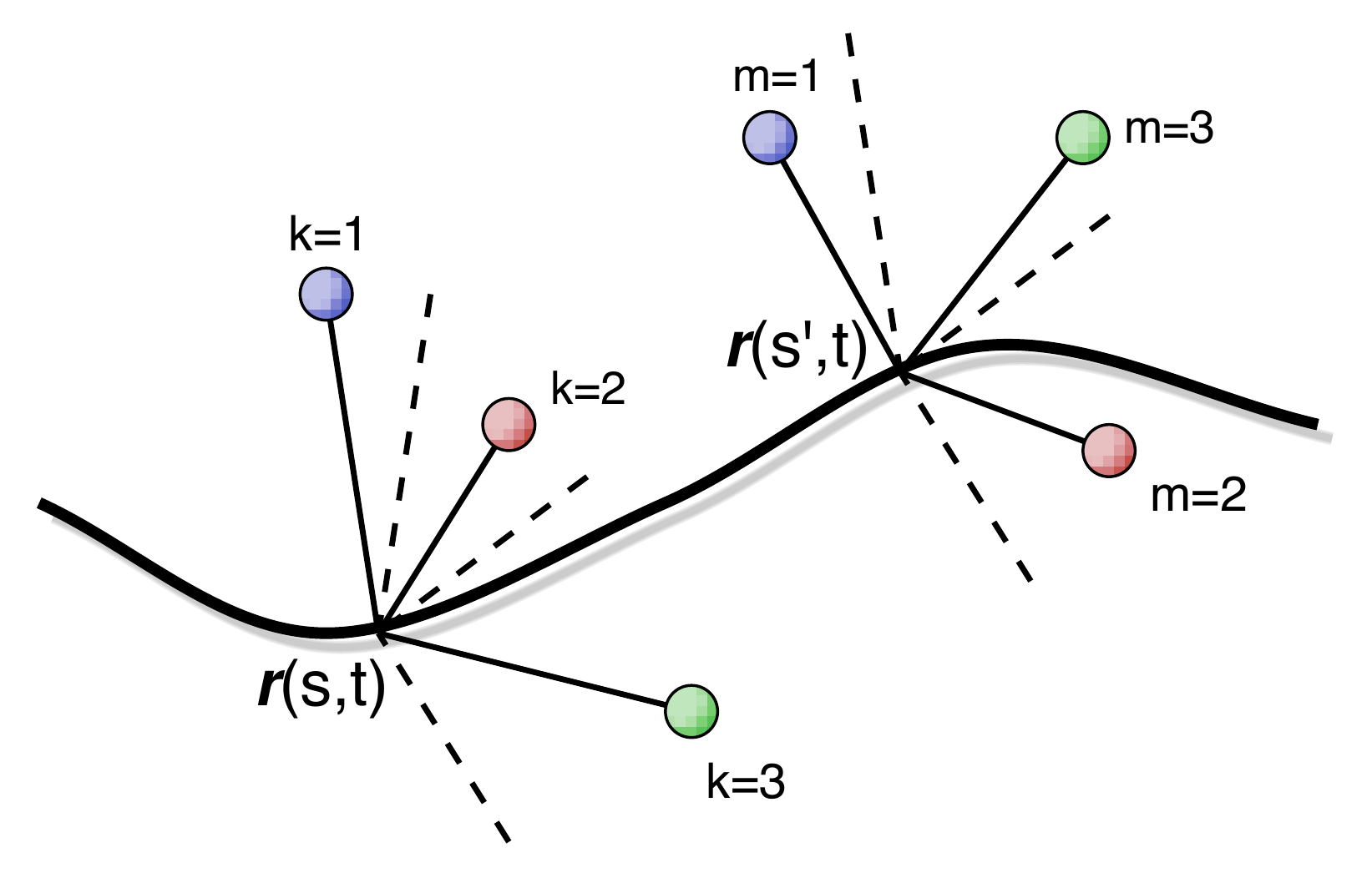}
 \caption{Rigid conformations of charges are distributed along a curve.}
 \label{model-fig}
 \end{figure}

The Skyrme model illustrates reduction of a principal bundle by its structure group as described in \cite{CaRaSh2000}.  Correspondingly, the \textit{molecular strand} demonstrates reduction of principal bundles by a subgroup of the structure group, which was the subject of \cite{CaRa2003}.  A molecular strand may be modeled as a flexible, elastic filament moving in $\mR^3$ with rigid charge conformations undergoing rigid rotations mounted along the filament's length, as shown in Figure \ref{model-fig}.  A full treatment of the molecular strand  was undertaken in \cite{ElGBHoPuRa2009}.

\paragraph{The geometry of the molecular strand.} The parameter space for the molecular strand is $X = I \times \mR$ where $I$ is an interval of fixed length and $\mathbb{R}$ represents time. The strand may be described by maps
\[
(\Lambda,\br):X \to SE(3)\cong SO(3)\times \mR^3,\quad x=(s,t)\mapsto \lp \Lambda(x), \br(x) \rp.
\]
Here $\br:X\to \mR^3$ describes the position of a point on the filament at a given time and $\Lambda:X\to SO(3)$ describes the rigid charge conformations along the filament at a given time.  These maps correspond to sections $\sigma$ of the principal bundle
\[
\pi_{X,P}: P := X \times SE(3) \to X,\quad\pi_{X,P}(x,\Lambda,\br):=x,
\]
by the relation
\[
\sig(x) = \lp x, \br(x),\Lambda(x)\rp.
\]

The Lagrangian $L:J^1P \to \mR$ of the molecular strand is taken to be left $SO(3)$-invariant as in \cite{ElGBHoPuRa2009}. Contrary to the case of the Skyrme model, the symmetry group of the theory \textit{does not} coincide with the structure group $SE(3)$ of the principal bundle $P$, but is a is a \textit{subgroup} of $SE(3)$. Thus, departing from the Skyrme model, $P$ is a principal $SO(3)$-bundle with the projection
\[
\pi_{\Sig,P}:P \to \Sig:= X \times \mR^3, \quad \pi_{\Sig,P}\lp x,\br, \Lambda\rp = \lp x, \Lambda^{-1}\br\rp.
\]
Now there are two bundle structures on $P$ given by $\pi_{X,P}$ and $\pi_{\Sig,P}$.  These two bundle structures induce a third, this time on $\Sig=X\times \mR^3$, given by
\[
\pi_{X,\Sig}: \Sig \to X,\quad \pi_{X,\Sig}(x,\rho)=x.
\]
The geometry is described by the commutative diagram.
\begin{diagram}
X\times SE(3)    &                        &  \rTo^{\boldsymbol{\pi_{X,P}}}  &                        &  X\\
                 & \rdTo_{\boldsymbol{\pi_{\Sig,P}}}   &                    & \ruTo_{\boldsymbol{\pi_{X,\Sig}}}   &   \\
                 &                        &   X\times \mR^3    &                        &   \\
\end{diagram}

\paragraph{The reduced bundle.} Since the symmetry group is a subgroup of the structure group, there is an additional part to the bundle isomorphism \eqref{isom_skyrme}. More precisely, a principal connection $\mathcal{A}$ on $\pi_{\Sigma,P}$ furnishes $J^1P/SO(3)$ with the vector bundle isomorphism
\[
J^1P/SO(3) \rightarrow J^1\Sig \oplus_\Sig L\lp \pi^*_{X,\Sig}TX, \Ad P\rp
\]
given by
\begin{equation}
\lsb T\sig\rsb_{SO(3)}\mapsto \left(T\lp \pi_{\Sig,P}\circ \sig\rp, \llsb \sig, \sig^*\cA\rrsb_{\mso(3)}\right).
\label{eq:MolBunIso}
\end{equation}
This isomorphism will be studied in detail in \S\ref{reduced_cov_conf_space}. Since, for the molecular strand, $\pi_{\Sig,P}$ is a trivial bundle, the adjoint bundle $\operatorname{Ad}P$ is also trivial and can be identified with $\Sigma\times\mathfrak{so}(3)\rightarrow\Sigma$ (projection on the first factor) via the isomorphism $\llsb(x,\Lambda,\br),\xi\rrsb_{\mathfrak{so}(3)}\mapsto (x,\Lambda^{-1}\br,\operatorname{Ad}_{\Lambda^{-1}}\xi)$. Using the Maurer-Cartan connection 
\[
\cA\lp v_x, v_{\br}, v_\Lambda\rp = v_\Lambda\Lambda^{-1},
\]
the isomorphism \eqref{eq:MolBunIso} reads
\[
\lsb \lp x, \br, \Lambda, \id, d\br, d\Lambda\rp\rsb_{SO(3)}\mapsto \lp x, \brho, d\brho, \Lambda^{-1}d\Lambda\rp = \lp x, \brho, \dot \brho dt + \brho' ds,  \om dt + \Om ds\rp
\]
where $\brho = \Lambda^{-1}\br$, $\om= \Lambda^{-1}\dot \Lambda$, $\Om = \Lambda^{-1}\Lambda'$, $(\dot{\phantom{o}})$ denotes differentiation with respect to $t$, and $(\,'\,)$ denotes differentiation with respect to $s$.

\paragraph{The reduced Euler-Lagrange equations.} The main result from \cite{CaRa2003} states that when $\pi_{X,P}:P \to X$ is a principal bundle and $G$ is a subgroup of the structure group, then the Euler-Lagrange equations on $P$ for a left $G$-invariant Lagrangian $L$ are equivalent to the Lagrange-Poincar\'e equations for the reduced Lagrangian $l$ on the reduced bundle $J^1P/G \cong J^1\Sig \oplus_\Sig L\lp \pi_{X,\Sig}^*TX,\Ad P\rp$.  The Lagrange-Poincar\'e equations read
\begin{equation}\label{eq:MolStr}
\left\{
\begin{array}{l}
\displaystyle\vspace{0.2cm}\dive^\cA\dede{l}{\bar\sig} - \ad^*_{\bar\sig}\dede{l}{\bar\sig}=0
,\\
\displaystyle\dive \dede{l}{d\brho} - \dede{l}{\brho} =0.
\end{array}
\right.
\end{equation}
The exact definitions of the differential operators and the functional derivatives appearing in these equations are studied in \S\ref{sec:geometric_constructions} and \S\ref{sec:LP_field}.
The right hand side of the first equation in \eqref{eq:MolStr} usually has an extra term associated with the curvature of the principal connection $\cA$.  This term is described in more detail below, but for the present case $\cA$ is flat, so the curvature term vanishes.  In local coordinates equations \eqref{eq:MolStr} read
\[
\left\{
\begin{array}{l}
\displaystyle\vspace{0.2cm}\lp \prt_t + \om \times \rp \dede{l}{\om} + \lp \prt_s + \Om\times\rp \dede{l}{\Om} =0,\\
\displaystyle\prt_t \dede{l}{\dot \brho} + \prt_s\dede{l}{\brho'} - \dede{l}{\brho} =0.
\end{array}
\right.
\]
These equations need to be augmented with an integrability condition to allow reconstruction.  This integrability/reconstruction condition is related to the curvature of $\cA$. For the molecular strand the required reconstruction condition is
\[
\prt_t \Om - \prt_s\om - \om \times \Om =0.
\]
These equations recover the results derived in \cite{ElGBHoPuRa2009} where the Lie algebra, $\mso(3)$ was identified with $\mR^3$ and therefore the adjoint actions became cross-products.  More details about the reconstruction condition can be found in \S\ref{sec:Reconstruction}.

\subsection{Fiber bundle reduction}\label{sec:fibre}

In \S\ref{sec:Principal} we introduced the ideas of \cite{CaRaSh2000} in the context of the classical Skyrme model.  Reduction of such `pure gauge' theories requires the introduction of certain geometric tools such as the adjoint bundle and jet bundles. In \S\ref{sec:Subgroup} we used the example of the molecular strand to review how principal bundle reduction can be extended to include reduction on principal bundles by a subgroup of the structure group as in \cite{CaRa2003}.  The purpose of this paper is to extend these ideas farther to include \textit{fiber bundle reduction}.

\paragraph{Lagrangian reduction in classical mechanics.} Consider \textit{classical Lagrangian reduction} from the perspective of reduction of variational principles.  A variational principle is formulated on a principal bundle $\pi:Q \to Q/G$ and a principal connection $\cA$ is introduced on $Q$.  The connection yields a bundle isomorphism
\[
(TQ)/G \rightarrow T\lp Q/G\rp \oplus_{Q/G} \Ad Q
\]
given by
\[
\lsb v_q\rsb \mapsto \lp T\pi\lp v_q\rp, \llsb q, \cA\lp v_q\rp \rrsb_{\mg}\rp.
\]
Thus, a curve $q(t)\in Q$ induces the two curves
\[
\rho(t):=\pi(q(t))\in Q/G\quad\text{and}\quad \quad \bar\sig(t) = \llsb q(t), \cA\lp \dot q(t)\rp \rrsb_{\mg}\in \Ad Q.
\]
Classical Lagrangian reduction states that the Euler-Lagrange equations on $Q$ with a $G$ invariant Lagrangian $L$ are equivalent to the Lagrange-Poincar\'e equations on $TQ/G \cong T\lp Q/G\rp \oplus_{Q/G} \Ad Q$ with reduced Lagrangian $l$.  The Lagrange-Poincar\'e equations read
\[
\left\{
\begin{array}{l}
\displaystyle\vspace{0.2cm}\DD{}{t}\dede{l}{\bar\sig} - \ad^*_{\bar \sig}\dede{l}{\bar\sig}=0
,\\
\displaystyle\dede{l}{\rho} - \DD{}{t}\dede{l}{\dot\rho} = \scp{\dede{l}{\bar\sig}}{{\bf i}_{\dot\rho}\tilde \cB},
\end{array}
\right.
\]
where $\tilde \cB$ is the reduced curvature form associated to the connection $\cA$ and $D/Dt$ denotes \ suitable covariant derivative.

\paragraph{Extension to field theories.} When this classical approach is generalized to field theories the natural structure to consider is the trivial fiber bundle
\[
\pi_{X,P}: P: = X\times Q\rightarrow X,\quad \pi_{X,P}(x,q):=x.
\]
Now $\Gamma\lp \pi_{X,P}\rp$, the space of sections of $\pi_{X,P}$, generalizes the space of curves in $Q$ and the principal bundle structure $\pi$ on $Q$ gives a natural principal bundle structure
\[
\pi_{\Sig,P}: P:=X\times Q \to \Sig :=X\times \lp Q/G\rp,\quad \pi_{\Sig, P}(x,q):=(x,\pi(q)).
\]
More generally, one would consider the following commutative diagram
\begin{diagram}
P &                       & \rTo^{\boldsymbol{\pi_{X,P}}} &                       & X \\
  & \rdTo_{\boldsymbol{\pi_{\Sig, P}}} &                  & \ruTo_{\boldsymbol{\pi_{X,\Sig}}}  &   \\
  &                       & \Sig             &                       &   \\
\end{diagram}
where $\pi_{X,P}:P\rightarrow X$ is {\em any} fiber bundle and $\pi_{\Sig, P}:P\rightarrow\Sigma$ is a principal bundle, whose group action preserves the fibers of $\pi_{X,P}$.  This situation arises, for example, in the isoperimetric problem and in image dynamics, as is outlined in \S\ref{sec:Applications}.

\paragraph{Goals of the paper.} Following the preceding discussion, a specification of properties emerges.  One would like to develop a framework for reduction that:
\begin{enumerate}
\item Captures the natural generalization of classical Lagrange-Poincar\'e reduction to field theories.
\item Reduces to classical Lagrange-Poincar\'e reduction as a particular case.
\item Reduces to the work of \cite{CaRaSh2000} and \cite{CaRa2003} in particular cases.
\end{enumerate}

In \S\ref{sec:geometric_constructions} some geometric tools that are necessary for performing reduction are introduced.  The relationship between two bundle structures on the same manifold $P$ is also studied. In
\S\ref{sec:LP_field}, the Lagrange-Poincar\'e field equations are developed and  a method for reconstruction is given. The Kelvin-Noether theorem associated to the Lagrange-Poincar\'e field equations is presented in  \S\ref{cons_law}. Finally, in \S\ref{sec:Applications}, the reduction tools developed earlier are applied to the isoperimetric problem and image dynamics.
Throughout the paper, constant reference to four particular cases is made in order to illustrate the theory: the \textit{unreduced case}, \textit{principal bundle reduction}, \textit{subgroup reduction}, and \textit{classical reduction}.

\section{Geometric constructions}\label{sec:geometric_constructions}

There are two main geometric constructions of interest.  The first is the interaction of two bundle structures, $\pi_{X,P}$ and $\pi_{\Sig,P}$ on $P$.  The second is the reduction of the jet bundle $J^1P$ by the structure group $G$.

\subsection{Geometric setting}
\label{sec:geometric_setting}

Consider a locally trivial fiber bundle $\pi_{X,P}:P\rightarrow X$. A \textit{section} of $\pi_{X,P}$ is a smooth map $\sigma:X\rightarrow P$ such that $\pi_{X,P}\circ\sigma=\id_X$. We denote by $\Gamma(\pi_{X,P})$ the set of all smooth sections of $\pi_{X,P}$.

\paragraph{Remark.}  It is necessary to introduce many fiber bundle projections during the development of the theory.  The notation indicates the source and target space, e.g.\ $\pi_{X,P}:P \to X$, where the first subscript denotes the base space and the second the total space.  The order of the subscripts allows one to write, for example,
\[
\pi_{X,Y}\circ \pi_{Y,Z} = \pi_{X,Z}.
\]

The \textit{first jet bundle} of $\pi_{X,P}$ is the affine bundle $\pi_{P,J^1P}:J^1P\rightarrow P$ whose fiber at $p$ is the affine space
\[
J^1_pP=\left\{\gamma_p\in L(T_xX,T_pP)\mid T_p\pi_{X,P}\circ\gamma_p=\id_{T_xX},\; x=\pi_{X,P}(p)\right\},
\]
where $L(T_xX,T_pP)$ denotes the space of linear maps $\gamma_p:T_xX\rightarrow T_pP$.  The first jet bundle is the natural generalization of the tangent bundle to the field theoretic context.  Therefore $J^1P$ plays the role of our {\it unreduced state space} in applications.  The manifold $J^1P$ may also be regarded as a locally trivial fiber bundle over $X$, that is, $\pi_{X, J^1P}:J^1P\rightarrow X$ with $\pi_{X, J^1P}(\gamma_p):=\pi_{X,P}(p)$. Given $\sigma\in\Gamma(\pi_{X,P})$, the \textit{first jet extension} of $\sigma$, defined by $j^1\sigma(x):=T_x\sigma$, is a section of $\pi_{X,J^1P}$.

Suppose there is a \textit{free and proper left\/} action $\Phi$ of a Lie group $G$ on $P$ such that
\begin{equation}\label{compatibility_condition}
\pi_{X,P}\circ\Phi_g=\pi_{X,P},\;\;\text{for all}\;\;g\in G.
\end{equation}
Equation \eqref{compatibility_condition} is equivalent to the assumption that the action of $G$ preserves the fibers of $\pi_{X,P}$. Since the action is free and proper, there exists a principal bundle $\pi_{\Sigma, P}:P\rightarrow\Sigma$, where $\Sigma:=P/G$.  Here $\Sig$ is the equivalent of {\it shape space} in applications. Since, by \eqref{compatibility_condition}, the projection $\pi_{X,P}$ is $G$-invariant, it induces a surjective submersion $\pi_{X,\Sigma}:\Sigma\rightarrow X$ via the relation
\begin{equation}\label{compatibility_condition2}
\pi_{X,\Sigma}\circ\pi_{\Sigma, P}=\pi_{X,P}.
\end{equation}
It is easily verify that if $\pi_{X,P}$ is proper then $\pi_{X,\Sigma}$ is also proper. 

More generally if $\pi_{X,P}$ is a locally trivial fiber bundle then $\pi_{X,\Sigma}$ is also a locally trivial fiber bundle.
To see this, take a fiber bundle chart $\psi:\pi_{X,P}^{-1}(U)\rightarrow U\times F$, where $U$ is an open subset of $X$ and the manifold $F$ is the model of the fiber.  By definition, $p_1\circ \psi=\pi_{X,P}$, where $p_1: U \times F \rightarrow U$ is the projection onto the first factor. Property \eqref{compatibility_condition} implies that $\pi_{X,P}^{-1}(U)$ is a $G$-invariant subset. Thus, the diffeomorphism $\psi$ bestows a well-defined $G$-action on $U\times F$ which turns out to be free, proper, and acts only on the component $F$.  The model fiber $F$ thereby attains a principal bundle structure $F\rightarrow F/G$ induced by (and depending upon) the chart $\psi$.
Since $\psi$ is an equivariant diffeomorphism, it drops to a diffeomorphism $\bar\psi:\pi_{X,P}^{-1}(U)/G\rightarrow U\times F/G$. Also, since $\pi_{X,P}^{-1}(U)/G=\pi_{X,\Sigma}^{-1}(U)$ and $p_1\circ\bar\psi=\pi_{X,\Sigma}$, the map $\bar\psi$ is a fiber bundle chart of $\pi_{X,\Sigma}:\Sigma\rightarrow X$.
For principal bundles, one needs to work with local sections, since a principal bundle does not have global sections unless it is trivial.

\paragraph{Remark.} There are now two different bundles $\pi_{X,P}:P\rightarrow X$ and $\pi_{\Sigma, P}:P\rightarrow\Sigma$ with the same total space $P$. In general, the associated vertical distributions do not coincide, although \eqref{compatibility_condition} provides the inclusion
\[
\operatorname{ker}\left(T_p\pi_{\Sigma, P}\right)\subset \operatorname{ker}\left(T_p\pi_{X,P}\right).
\]
Thus, it is possible to associate two different jet bundles to $P$. Throughout this paper only the first jet bundle of interest is $\pi_{X,P}: P\rightarrow X$ and hence there is no ambiguity in the notation $J^1P$.

\medskip

Lagrangian field theories are described by a Lagrangian density $\mathcal{L}:J^1P\rightarrow \Lambda^{n+1}X$ defined on the first jet bundle. Here $\Lambda^{n+1}X$ denotes the bundle of $(n+1)$-forms on $X$, where $n+1=\operatorname{dim}X$. In this context the $G$-action on $P$, lifted to $J^1P$, should be interpreted as a \textit{symmetry} of the Lagrangian density. The associated reduction process, described in the next section, is called the \textit{covariant Lagrange-Poincar\'e reduction}.

\paragraph{Particular cases.} Various previous theories may be identified as particular cases of the geometric setting developed in this paper. These examples will be referred to throughout the paper and serve to illustrate the ideas introduced in more familiar contexts while demonstrating how the objective of capturing previous theories in this new context is fulfilled.
\begin{itemize}

\item[{\bf i}] If $\Sigma=P$, that is, $G=\{1\}$, there are no symmetries. The principal bundle structure disappears and the geometric setting for \textit{covariant Lagrangian field theory}, referred to as the {\em unreduced case},  emerges.  The commutative diagram that describes this case is
\begin{diagram}
P &                                     & \rTo^{\boldsymbol{\pi_{X,P}}} &                                  &  X      \\
  &  \rdTo_{\boldsymbol{\pi_{\Sig,P} = \id_P}}  &                               & \ruTo_{\boldsymbol{\pi_{X,\Sig} =\pi_{X,P}}}       &         \\
  &                                                                                                                            &                 \Sig=P                                                                    &                                                                                                             &                  \\
\end{diagram}
where $\pi_{X,P}$ is a fiber bundle.

\item[{\bf ii}] Assume that the configuration space $\pi_{X,P}: P \to X$ is itself a principal $G$-bundle and $G$ is also the symmetry group. Then $\Sigma=X$ and $\pi_{X, \Sigma}$ is the identity map. This recovers  the geometric setting for \textit{covariant Euler-Poincar\'e reduction}, or {\em principal bundle reduction} in \cite{CaRaSh2000} which is used to study, for example, the Skyrme model.  The commutative diagram that describes this case is
\begin{diagram}
P &                                     & \rTo^{\boldsymbol{\pi_{X,P}}} &                                  &  X      \\
  &  \rdTo_{\boldsymbol{\pi_{\Sig,P} = \pi_{X,P}}}  &                               & \ruTo_{\boldsymbol{\pi_{X,\Sig} =\id_X}}       &         \\
  &                                                                                                                            &                 \Sig=X                                                                    &                                                                                                             &                  \\
\end{diagram}
where $\pi_{X,P}$ is a principal bundle.
\item[{\bf iii}]  If $\pi_{X,P}:P\rightarrow X$ is a principal bundle whose structure group $H$ contains the group of symmetries $G$ as a \textit{subgroup} one recovers the formulation in \cite{CaRa2003}.  This is the geometric setting for the formulation of the molecular strand from \cite{ElGBHoPuRa2009} and is referred to as \textit{subgroup covariant Lagrange-Poincar\'e reduction} or simply {\em subgroup reduction}.  The commutative diagram that describes this case is
\begin{diagram}
P &                                     & \rTo^{\boldsymbol{\pi_{X,P}}} &                                  &  X      \\
  &  \rdTo_{\boldsymbol{\pi_{\Sig,P}}}  &                               & \ruTo_{\boldsymbol{\pi_{X,\Sig}}}&         \\
  &                                                                                                                            &                 \Sig                                                                              &                                                                                                             &                  \\
\end{diagram}
where $\pi_{X,P}$ and $\pi_{\Sig, P}$ are respectively $G $ and $H $-principal bundles.

\item[{\bf iv}] If $X=\mathbb{R}$, $\Sigma=\mathbb{R} \times M$, and $P=\mathbb{R}\times Q$, where $\pi_{M, Q}:Q\rightarrow M$ is a $G$-principal bundle, the geometric setting for Lagrangian reduction in classical mechanics, known as {\em classical reduction}, becomes apparent. Here $Q$ plays the role of the configuration space. There are two well-known particular cases: If $Q=G$ (thus $M=\{m\}$) the geometric setting for \textit{Euler-Poincar\'e reduction} surfaces (this is also a particular case of \textbf{ii}), where the configuration space coincides with the group of symmetries; if $G=\{1\}$ (thus $Q=M$) there are no symmetries and we reacquire the geometric setting for unreduced classical Lagrangian mechanics (this is also a particular case of \textbf{iii}).  The commutative diagram that describes this case is
\begin{diagram}
P=\mR\times Q &                            & \rTo^{\boldsymbol{\pi_{X,P}=p_1}} &                           &  X = \mR      \\
  &  \rdTo_{\boldsymbol{\id_\mR\times \pi_{M,Q}}}  &                               & \ruTo_{\boldsymbol{\pi_{X,\Sig}= p_1}}&         \\
  &                                                                                                                            &                 \Sig = \mR\times M                                                                             &                                                                                                             &                  \\
\end{diagram}
where $\pi_{M,Q}$ is a principal bundle.
\end{itemize}

\medskip

\paragraph{Adjoint bundle.} The \textit{adjoint bundle} associated with the principal bundle $\pi_{\Sigma, P}:P\rightarrow \Sigma$ is a \textit{vector bundle} $\pi_{\Sigma, \operatorname{Ad}P}:\operatorname{Ad}P\rightarrow\Sigma$.   The total space $\operatorname{Ad}P$ is the quotient space $P\times_G\mathfrak{g}$ relative to the following left action of $G$ on $P\times\mathfrak{g}$:
\[
(p,\xi)\mapsto (\Phi_g(p),\operatorname{Ad}_g\xi).
\]
Elements in the adjoint bundle are equivalence classes $\llsb p,\xi\rrsb_{\mg}$ and the projection is described by
\[
\pi_{\Sigma,\operatorname{Ad}P}\left(\llsb p,\xi\rrsb_{\mg}\right)=\pi_{\Sigma, P}(p).
\]
The adjoint bundle is, in fact, a Lie algebra bundle.  That is, each fiber 
$(\operatorname{Ad} P)_s$, $s \in \Sigma$, has a natural Lie bracket 
\[
\big[ \llsb p, \xi\rrsb_{\mg}, \llsb p, \eta\rrsb_{\mg} \big]_s : = \llsb p, \lsb \xi, 
\eta\rsb\rrsb_{\mg},
\]
where $\pi_{\Sigma , P}(p) = s $, $\xi, \eta\in \mathfrak{g}$  and these Lie brackets depend smoothly on the base variable $s \in \Sigma$.

This Lie algebra bundle structure enables the introduction of a wedge product.  For $1$-forms this wedge product $\wedge : \Omega^1(\Sigma,\operatorname{Ad}P)\times  \Omega^1(\Sigma,\operatorname{Ad}P)\rightarrow  \Omega^2(\Sigma,\operatorname{Ad}P)$ is defined by
\begin{equation}\label{wedge_bundle_valued}(\alpha\wedge\beta)(u_s,v_s):= [\alpha(u_s),\beta(v_s)]_s - [\alpha(v_s),\beta(u_s)]_s,
\end{equation}
where $s \in \Sigma $ and $u_s, v_s \in T_s \Sigma$.

The different equivalence classes are interpreted as different representations of the dynamics.  For given $s \in \Sigma$ and $p\in\pi^{-1}_{\Sig, P}\lp s\rp$ one can define a $p$-dependent Lie algebra isomorphism $\chi_p:(\operatorname{Ad}P)_s\rightarrow \mathfrak{g}$ by
\begin{equation}
\chi_p\left(\llsb q,\eta\rrsb_{\mg}\right)=\xi,
\label{eq:Rep}
\end{equation}
where $\xi\in\mathfrak{g}$ is such that $\llsb p,\xi\rrsb_{\mg}=\llsb q,\eta\rrsb_{\mg}$.

The choice of $p \in \pi^{-1}_{\Sig, P}\lp s\rp$ determines the representation of the dynamics.  Thus by altering $p$ we use $\chi_p$ to give the dynamics in the {\it convective} or the {\it spatial} representation.

A \textit{connection} $\mathcal{A}$ on the principal bundle $\pi_{\Sigma, P}:P\rightarrow \Sigma$  is a one-form $\mathcal{A}\in\Omega^1(P,\mathfrak{g})$ such that
\[
\Phi_g^*\mathcal{A}=\operatorname{Ad}_g\circ\mathcal{A}\quad\text{and}\quad\mathcal{A}(\xi_P(p))=\xi,\quad\text{where}\quad \left.\xi_P(p):=\frac{d}{dt}\right|_{t=0}\Phi_{\operatorname{exp}(t\xi)}(p)
\]
is the \textit{infinitesimal generator} associated to the Lie algebra element $\xi\in\mathfrak{g}$. The \textit{horizontal distribution} associated to $\mathcal{A}$ is the subbundle $HP\subset TP$ defined by
\[
H_pP:=\operatorname{ker}\left(\mathcal{A}(p)\right).
\]
The horizontal distribution is complementary to the vertical distribution $V_pP:=\operatorname{ker}\left(T_p\pi_{\Sigma P}\right)$ and consequently $TP$ decomposes as $T_pP=H_pP\oplus V_pP$. The connection $\mathcal{A}$ defines the \textit{horizontal lift} operator $\operatorname{Hor}^\mathcal{A}_p:T_s\Sigma\rightarrow H_pP$ according to
\[
\operatorname{Hor}^\mathcal{A}_p(v_s)\in H_pP\quad \text{and}\quad T_s\pi_{\Sigma P}\circ\operatorname{Hor}^\mathcal{A}_p=\id_{T_s\Sigma}
\]
where $s \in \Sigma$, $p \in \pi^{-1}_{\Sigma, P}\lp s\rp$ and $v_s \in T_s\Sigma$.

The connection $\mathcal{A}$ also induces a \textit{covariant derivative} on the adjoint bundle
\begin{equation}\label{adjoint_cov_der}
\nabla^\mathcal{A}:\Gamma(\pi_{\Sigma, \operatorname{Ad}P})\rightarrow\Gamma\left(\pi_{\Sigma, L(T\Sigma,\operatorname{Ad}P)}\right)
\end{equation}
given by
\begin{equation}\label{adjoint_cov_der_formula}\nabla^\mathcal{A}_{v_s}\sigma(s)=\left.\frac{D^\mathcal{A}}{Dt}\right|_{t=0}\sigma(c(t))=\llsb p(s),\mathbf{d}\xi(v_s)-[\mathcal{A}(Tp(v_s)),\xi(s)]\rrsb_{\mg},
\end{equation}
where $\xi:\Sigma\rightarrow \mathfrak{g}$ and $p:\Sigma\rightarrow P$ are such that $\sigma(s)=\llsb p(s),\xi(s)\rrsb_{\mg}$ and $c(t)$ is a curve in $\Sigma$ such that $\dot{c}(0)=v_s$ (see \cite{CeMaRa2001}, Lemma 2.3.4). The covariant derivative $\nabla^\mathcal{A}$ also has an interpretation as a bilinear map
\[
\nabla^\mathcal{A}:\mathfrak{X}(\Sigma)\times \Gamma\left(\pi_{\Sigma, \operatorname{Ad}P}\right)\rightarrow \Gamma\left(\pi_{\Sigma, \operatorname{Ad}P}\right),\;\; (X,\sigma)\mapsto \nabla_X^\mathcal{A}\sigma.
\]

\subsection{Reduced covariant configuration space}\label{reduced_cov_conf_space}

The free and proper action $\Phi:G\times P\rightarrow P$ induces a free and proper action $\Phi^1:G\times J^1P\rightarrow J^1P$ defined by
\begin{equation}\label{induced_action_on_jets}
\Phi^1_g(\gamma_p):=T\Phi_g\circ\gamma_p, \qquad \gamma_p \in J^1_p P.
\end{equation}
Note that this action preserves $J^1P$ since, by \eqref{compatibility_condition},
\[
T\pi_{XP}\circ \Phi^1_g(\gamma_p)=T\pi_{XP}\circ T\Phi_g\circ\gamma_p=T\pi_{XP}\circ\gamma_p=\id_{T_xX}.
\]
Thus it is valid to consider the quotient manifold $J^1P/G\ni [\gamma_p]_G$.

\paragraph{Remark.} Recall that $J^1P$ denotes the first jet bundle of $P$ as a fiber bundle over $X$ and not as a principal bundle over $\Sigma$.

\medskip 

The connection $\mathcal{A}$ on the principal bundle $\pi_{\Sigma P}:P\rightarrow\Sigma$ introduces the smooth map $\beta_\mathcal{A}$, which is defined by
\begin{equation}\label{definition_beta}
\beta_\mathcal{A}:J^1P/G\rightarrow J^1\Sigma\oplus_\Sigma L\left(\pi_{X,\Sigma}^*TX,\operatorname{Ad}P\right),\;\;\beta_\mathcal{A}\left([\gamma_p]_G\right):=(T\pi_{\Sigma, P}\circ\gamma_p,\llsb p,\mathcal{A}(\gamma_p(\_\,))\rrsb_{\mg}),
\end{equation}
where
\[
\pi_{\Sigma,L\left(\pi_{X,\Sigma}^*TX,\operatorname{Ad}P\right)}:L(\pi_{X,\Sigma}^*TX,\operatorname{Ad}P)\rightarrow\Sigma
\]
is the vector bundle whose fiber at $s\in\Sigma$ is $L(T_{\pi_{X, \Sigma}(s)}X,\operatorname{Ad}P_s)$.

The map $\beta_\mathcal{A}$ is an diffeomorphism, the inverse being given by
\[
\beta_\mathcal{A}^{-1}:J^1\Sigma\oplus_\Sigma L\left(\pi_{X,\Sigma}^*TX,\operatorname{Ad}P\right)\rightarrow J^1P/G,\;\;\beta_\mathcal{A}^{-1}(\delta_s,l_s)=\left[\operatorname{Hor}^\mathcal{A}_p\circ\,\delta_s+\zeta_p\circ l_s\right]_G,
\]
where $p\in P$ is such that $\pi_{\Sigma, P}(p)=s$ and 
\[
\zeta_p:(\operatorname{Ad}P)_s\rightarrow V_pP
\]
is defined by $\zeta_p\left([q,\eta]_G\right)=\xi_P(p)$ where $\xi\in\mathfrak{g}$ is such that $\llsb q,\eta\rrsb_\mg=\llsb p,\xi\rrsb_\mg$. Note that $\operatorname{Hor}^\mathcal{A}_p\circ\,\delta_s+\zeta_p\circ l_s \in  J^1_pP$ and that its equivalence class does not depend on which $p$ is chosen. Also, note that the diffeomorphism $\beta_\mathcal{A}$ endows the manifold $J^1P/G$ with the structure of an affine bundle over $\Sigma$.

\medskip

The isomorphism $\beta_\mathcal{A}$ is interpreted as follows in the four particular cases:
\begin{itemize}
\item[{\bf i}] Here $G=\{1\}$ thus the principal bundle structure disappears. The bundle isomorphism is the identity on $J^1P$.
\item[{\bf ii}] Here $\Sigma=X$, thus $\beta_\mathcal{A}$ is a bundle map over $X$ and we have
\[
\beta_\mathcal{A}:J^1P/G\rightarrow L(TX,\operatorname{Ad}P),\quad \beta_\mathcal{A}([\gamma_p])=\llsb p,\mathcal{A}(\gamma_p(\_\,))\rrsb_{\mg}
\]
and we recover the isomorphism used in the Euler-Poincar\'e reduction, see formula (2.5) in \cite{CaRaSh2000}.
\item[{\bf iii}] The isomorphism used in the subgroup Lagrange-Poincar\'e reduction reemerges, see Proposition 3 in \cite{CaRa2003}.

\item[{\bf iv}] Since $X=\mathbb{R}$ and $P=\mathbb{R}\times Q$, the jet bundle $J^1P$ may be identified with $\mathbb{R}\times TQ$. Thus, $J^1P/G\simeq \mathbb{R}\times (TQ/G)$. Similarly, $J^1\Sigma$ may be identified with $\mathbb{R}\times TM$ and $\operatorname{Ad}P$ with $\mathbb{R}\times\operatorname{Ad}Q$. The bundle $J^1\Sigma\oplus_\Sigma L\left(\pi_{X,\Sigma}^*TX,\operatorname{Ad}P\right)$ can thus be identified with $\mathbb{R}\times \left(TM\oplus_M\operatorname{Ad}Q\right)$. A connection $\gamma$ on $Q$ naturally induces a connection $\mathcal{A}$ on $P$ and the bundle map $\beta_\mathcal{A}$ reads $\beta_\mathcal{A}: \mathbb{R}\times (TQ/G)\rightarrow\mathbb{R}\times \left(TM\oplus_M\operatorname{Ad}Q\right)$,
\begin{equation}\label{beta_A_special}
\beta_\mathcal{A}\left(t,[v_q]_G\right)=\left(t,T\pi_{M,Q}(v_q),\llsb q,\gamma(v_q)\rrsb_{\mg}\right).
\end{equation}
Therefore the usual connection dependent isomorphism $TQ/G\simeq TM\oplus_M\operatorname{Ad}Q$ used in classical Lagrangian reduction is recovered, as in \cite{CeMaRa2001}.

\end{itemize}

\section{Lagrange-Poincar\'e field equations}\label{sec:LP_field}

Consider a $G$-invariant Lagrangian density
$\mathcal{L}:J^1P\rightarrow \Lambda^{n+1}X$. For simplicity, suppose that $X$ is orientable and fix a volume form $\mu$ on $X$. The Lagrangian density may thereby be expressed as $\mathcal{L}=L\mu$, where $L:J^1P\rightarrow\mathbb{R}$.

Let $U\subset X$ be an open subset whose closure $\bar{U}$ is compact. Recall that a section $\sigma:\bar{U}\rightarrow P$ of $\pi_{X,P}$ is, by definition, smooth if  for every point $x \in \bar{U}$ there is an open neighborhood $U_x$ of $x $ and a smooth section $\sigma_x : U_x \rightarrow P$ extending $\sigma$. A \textit{critical section} of the variational problem defined by $\mathcal{L}$ is defined as a smooth local section $\sigma:\bar U\rightarrow P$ of $\pi_{X,P}$ that satisfies 
\[
\left.\frac{d}{d\varepsilon}\right|_{\varepsilon=0}\int_U\mathcal{L}(j^1\sigma_\varepsilon)=0,
\]
for all smooth \textit{variations} $\sigma_\varepsilon : \bar{U} \rightarrow P$ such that $\sigma_0=\sigma$ and $\sigma_\varepsilon|_{\partial U}=\sigma|_{\partial U}$. Since 
\[
\delta\sigma(x):=\left.\frac{d}{d\varepsilon}\right|_{\varepsilon=0}\sigma_\varepsilon(x) \in V_{\sigma(x)}P: =\operatorname{ker}(T_{\sigma(x)}\pi_{X,P})\quad\text{and}\quad\delta\sigma|_{\partial U}=0,
\]
one may assume without loss of generality that $\sigma_\varepsilon = \phi_\varepsilon\circ\sigma$, where $\phi_\varepsilon$ is the flow of a vertical (with respect to the bundle structure $\pi_{X,P}$) vector field $V\in\mathfrak{X}^V(P)$ such that $V(\sigma(x))=0$ for all $x\in\partial U$.  The smooth Tietze extension theorem facilitates $V $'s definition over the whole manifold $P $, but values of $V $ outside $\sigma(\bar{U}) $ will not play any role in any subsequent consideration. Note that 
$\delta \sigma(x) = V ( \sigma(x))$ for all $x \in \bar{U} $. Consequently,
\[
\delta j^1\sigma(x):=\left.\frac{d}{d\varepsilon}\right|_{\varepsilon=0}j^1\sigma_\varepsilon(x)=j^1V(j^1\sigma(x))\in V_{j^1\sigma(x)}J^1P=\operatorname{ker}(T_{j^1\sigma(x)}\pi_{X,J^1P}),
\]
where $V\in \mathfrak{X}^V(P)\mapsto j^1V\in \mathfrak{X}^V(J^1P)$ is the $1$-\textit{jet lift of vector fields}.  Thus, $\sig$ is a critical section of the variational problem defined by $\cL$ if
\[
0=\left.\frac{d}{d\varepsilon}\right|_{\varepsilon=0}\int_UL(j^1\sigma_\varepsilon(x))\mu=\int_U\left\langle\frac{\delta L}{\delta\sigma}(x),j^1V(j^1\sigma(x))\right\rangle\mu,
\]
where $\lp \delta L/\delta\sigma\rp (x) \in V ^\ast_{j ^1 \sigma(x)} J ^1P$ is the differential along $j^1\sigma$.  That is,
\[
\left\langle\frac{\delta L}{\delta \sigma}(x),Z(j^1\sigma(x))\right\rangle=\mathbf{d}L(j^1\sigma(x))(Z(j^1\sigma(x)))
\]
for arbitrary vector fields $Z\in\mathfrak{X}^V(J^1P)$ that are vertical with respect to $\pi_{X,J^1P}$. Denoting by $\mathcal{EL}(L)$ the bundle morphism $\mathcal{EL}(L):J^1P\rightarrow V^*P$ defined by the condition
\[
\int_U\left\langle\mathcal{EL}(L)(j^1\sigma(x)),V(\sigma(x))\right\rangle\mu=\int_U\left\langle\frac{\delta L}{\delta\sigma}(x),j^1V(j^1\sigma(x))\right\rangle\mu,\;\;\text{for all}\;\; V\in\mathfrak{X}^V(P),
\]
the covariant Euler-Lagrange equations can be written intrinsically as
\[
\mathcal{EL}(L)=0.
\]
Here $\mathcal{EL}$ is represented locally by
\[
\mathcal{EL}(L)=\left[\frac{\partial L}{\partial y^\alpha}(j^1\sigma)-\frac{\partial}{\partial x^i}\left(\frac{\partial L}{\partial v^\alpha_i}(j^1\sigma)\right)\right]{\rm d}y^\alpha,
\]
where $\mathcal{L}=L(x^i,y^\alpha,v^\alpha_i)d^{n+1}x$.
Thus, in coordinates, the covariant Euler-Lagrange equations take the standard form
\begin{equation}\label{EL_local}
\frac{\partial L}{\partial y^\alpha}(j^1\sigma)-\frac{\partial}{\partial x^i}\left(\frac{\partial L}{\partial v^\alpha_i}(j^1\sigma)\right)=0.
\end{equation}
These equations may be written globally by using a connection on the affine bundle $\pi_{P,J^1P}:J^1P\rightarrow P$; this point of view will be used at the reduced level.

By $G$-invariance, $L$ induces the \textit{reduced Lagrangian} $l:J^1P/G\rightarrow \mathbb{R}$. Fixing a connection $\mathcal{A}$ on the principal bundle $\pi_{\Sigma, P}:P\rightarrow\Sigma$ brings in the bundle isomorphism $\beta_\mathcal{A}$, thereby permitting the definition of the reduced Lagrangian $l$ on $J^1\Sigma\oplus_\Sigma L\left(\pi_{X,\Sigma}^*TX,\operatorname{Ad}P\right)$.

A section $\sigma\in\Gamma(\pi_{X,P})$ of the configuration bundle induces a section
\[
\rho:=\pi_{\Sigma,P}\circ\sigma\in\Gamma(\pi_{X,\Sigma})
\]
by \eqref{compatibility_condition2}. The \textit{reduced section} is defined as
\begin{equation}\label{def_sigma_bar}\bar\sigma:=\llsb \sigma,\sigma^*\mathcal{A} \rrsb_{\mg}\in \Gamma\left(\pi_{X,L\left(\pi^*_{X,\Sigma}TX,\operatorname{Ad}P\right)}\right).
\end{equation}
Thus,
\[
(j^1\rho,\bar\sigma)=\beta_\mathcal{A}\left([j^1\sigma]_G\right):X\rightarrow J^1\Sigma\oplus_\Sigma L\left(\pi^*_{X,\Sigma}TX,\operatorname{Ad}P\right).
\]
The two components are not independent since $\rho$ can be obtained from $\bar\sigma$; explicitly,
\[
\rho=\pi_{\Sigma,L\left(\pi^*_{X,\Sigma}TX,\operatorname{Ad}P\right)}\circ\bar \sigma.
\]
Note that $(j^1\rho,\bar\sigma)$ is a section of the bundle $J^1\Sigma\oplus_\Sigma L\left(\pi^*_{X,\Sigma}TX,\operatorname{Ad}P\right)$ viewed as a fiber bundle over $X$, and not as an affine bundled over $\Sigma$. 
These definitions and the $G$-invariance of $\mathcal{L}$ (and hence of $L $) yield
\begin{equation}\label{three_Lagrangians}\mathcal{L}(j ^1 \sigma) = L(j ^1 \sigma) \mu= l(j^1\rho,\bar\sigma) \mu
\end{equation}
for any $\sigma \in \Gamma(\pi_{X, P}) $.

The previous considerations hold without changes when $\sigma$ is a local section $\sigma:\bar{U}\rightarrow P$.

The fact that $J^1\Sigma\oplus_\Sigma L\left(\pi^*_{X,\Sigma}TX,\operatorname{Ad}P\right)$ is a locally trivial fiber bundle over $X$ follows from the following observation: $G$ acts on the locally trivial fiber bundle $\pi_{X,J^1P}:J^1P\rightarrow X$ by a free and proper action $\Phi^1$, such that $\pi_{X,J^1P}\circ\Phi^1_g=\pi_{X,J^1P}$. Therefore, by the argument used in \S\ref{sec:geometric_setting}, $J^1P/G\rightarrow X$ is a locally fiber bundle. Thus, the isomorphism $\beta_\mathcal{A}$ ensures that $J^1\Sigma\oplus_\Sigma L\left(\pi^*_{X,\Sigma}TX,\operatorname{Ad}P\right)$ is a locally trivial fiber bundle over $X$.

\subsection{Reduced variations}

Using the bundle isomorphism $\beta_\mathcal{A}$, the variation of the action defined by $\mathcal{L}$ gives
\[
\left.\frac{d}{d\varepsilon}\right|_{\varepsilon=0}\int_U\mathcal{L}(j^1\sigma_\varepsilon)=\left.\frac{d}{d\varepsilon}\right|_{\varepsilon=0}\int_Ul(j^1\rho_\varepsilon,\bar\sigma_\varepsilon)\mu=0, 
\]
where $U \subset X $ is an open subset and $\sigma _\varepsilon: \bar{U} \rightarrow P $ is a smooth variation of the smooth section $\sigma:\bar{U} \rightarrow P $.

A covariant derivative on the locally trivial fiber bundle $\pi_{X,\operatorname{Ad}P}:=\pi_{X,\Sigma}\circ\pi_{\Sigma,\operatorname{Ad}P}:\operatorname{Ad}P\rightarrow X$ is required to compute the reduced variations. Recall the following general construction.

\paragraph{General constructions.} Let $\tau:E\rightarrow\Sigma$ a vector bundle endowed with a covariant derivative $\nabla$. Recall that $\nabla $ induces a covariant exterior derivative $\mathbf{d}^\nabla: \Omega^k(\Sigma, E) \rightarrow \Omega^{k+1}(\Sigma, E)$ whose formula is a direct adaptation of the standard Cartan formula for $k $-forms on a manifold, by replacing all directional derivatives by covariant derivatives relative to $\nabla$. In particular, for one-forms
\begin{equation}\label{covariant_nabla_exterior}\mathbf{d}^\nabla \alpha(U, V) = \nabla _U( \alpha(V)) - \nabla _V( \alpha(U)) - \alpha([U,V]),\end{equation}
where $\alpha\in \Omega^1(\Sigma , E) $ and $U, V \in \mathfrak{X}( \Sigma)$.

Let $X$ be an arbitrary manifold and $f:X\rightarrow E$ a smooth function. Define the $\nabla$-\textit{derivative} of $f$  by
\begin{equation}\label{nabla_tilde_general_def}
\tilde{\nabla}_{v_x}f(x):=\left.\frac{D^\nabla}{Dt}\right|_{t=0}f(c(t))\in E_{\tau(f(x))},
\end{equation}
where $c(t)\in X$ is a curve such that $\dot c(0)=v_x$ and $D^\nabla/ Dt$ is the usual covariant time derivative associated to $\nabla$ of the curve $t \mapsto f(c(t))$ in $E $. Note that $\tilde\nabla f(x)\in L\left(T_xX,E_{\tau(f(x))}\right)$, and when $U\in\mathfrak{X}(X)$, the derivative $\tilde{\nabla}_Uf$ is a function on $X$ taking values in $E$. 

The considerations below require an exterior covariant derivative of forms on $X$ with values in $E$. To make sense of this, assume that there is a smooth map $\pi: \Sigma \rightarrow X$. Recall that $\tau^k: L^k_a( \pi ^\ast TX , E) \rightarrow \Sigma$ is the vector bundle whose fiber at $s\in \Sigma$ is $L^k_a( \pi ^\ast TX , E)_s = L^k_a( T_{\pi(s)}X , E_s)$, the $k $-linear antisymmetric maps from $T_{\pi(s)}X$ to $E_s$. Define the $E $-valued $k$-forms on $X$ by
\[
\Omega^k_ \pi(X, E) : = \Gamma \left(\pi_{X, L^k_a(\pi ^\ast TX , E)} \right),
\]
where $\pi_{X, L^k_a(\pi ^\ast TX , E)} : = \pi\circ \tau^k:L^k_a(\pi ^\ast TX , E) \rightarrow X$. Note that this is not a vector bundle and thus $\Omega^k_ \pi(X, E)$ are not the usual vector bundle valued $k $-forms on $X$. In fact, $\Omega^k_ \pi(X, E)$ is \textit{not} even a vector space. In spite of this, there is a derivation, analogous to the usual exterior covariant differentiation \eqref{covariant_nabla_exterior} on forms. While the definition of this operator holds for general elements in $\Omega^k_ \pi(X, E) $ and is again based on Cartan's classical formula, only the definition for one-forms is needed:
\begin{equation}\label{tilde_covariant_nabla_exterior}\mathbf{d}^{\tilde{\nabla}} \alpha(U, V) = \tilde{\nabla}_U( \alpha(V)) - \tilde{\nabla} _V( \alpha(U)) - \alpha([U,V]),\end{equation}
where $\alpha \in \Omega^1_ \pi(X, E) $ and $U, V \in \mathfrak{X}(X) $. Note that $\alpha(U), \alpha(V):X \rightarrow E$, hence \eqref{nabla_tilde_general_def} is valid, also note that $\mathbf{d}^{\tilde{\nabla}} \alpha \in 
\Omega^2_ \pi(X, E)$.

Since $L^0_a(\pi^\ast TX, E ) = E$ by definition, $\tau^0 = \tau: E \rightarrow \Sigma$ and thus $\Omega^0_ \pi(X, E) = \Gamma( \pi_{X, E}) \subset C ^{\infty}(X, E)$, $\pi_{X,E} : = \pi\circ \tau^0$. Therefore the operator $\mathbf{d}^{\tilde{\nabla}}$ on $\Omega^0_ \pi(X, E)$ coincides with $\tilde{\nabla} $ as defined in \eqref{nabla_tilde_general_def}.
\medskip

\paragraph{Covariant derivatives.} Returning to the case at hand, the general construction specializes to the covariant derivative $\nabla^ \mathcal{A}$ on the vector bundle $\operatorname{Ad}P \rightarrow \Sigma$. Thus if $\xi\in\Gamma\left(\pi_{X,\operatorname{Ad}P}\right)$ and $u_x\in T_xX$, the previous definition, \ref{nabla_tilde_general_def}, gives the $\nabla^ \mathcal{A}$-derivative of $\xi$ by
\[
\tilde\nabla^\mathcal{A}_{u_x}\xi(x):=\left.\frac{D^\mathcal{A}}{Dt}\right|_{t=0}\xi(c(t)),
\]
where $c(t)$ is a curve in $X$ such that $\dot{c}(0)=u_x$. Writing $\xi(x)=\llsb p(x),\zeta(x)\rrsb_{\mg}$ yields
\begin{equation}\label{covariant_der_big}\tilde\nabla^\mathcal{A}_{u_x}\xi(x)=\llsb p(x),\mathbf{d}\zeta(u_x)-[\mathcal{A}(T_xp(u_x)),\zeta(x)]\rrsb_{\mg}
\end{equation}
(see \cite{CeMaRa2001}, Lemma 2.3.4).
Note that the $\nabla^\mathcal{A}$-derivative is a map
\[
\tilde\nabla^\mathcal{A}: \Gamma\left(\pi_{X,\operatorname{Ad}P}\right)\rightarrow\Gamma\left(\pi_{X,L(\pi^*_{X,\Sig}TX,\operatorname{Ad}P)}\right),
\]
not to be confused with \eqref{adjoint_cov_der}, and it can be interpreted as a map
\[
\tilde\nabla^\mathcal{A}: \mathfrak{X}(X)\times \Gamma\left(\pi_{X,\operatorname{Ad}P}\right)\rightarrow\Gamma\left(\pi_{X,\operatorname{Ad}P}\right).
\]
Note also that $\tilde\nabla^\mathcal{A}\xi$, $\tilde\nabla^\mathcal{A}_X\xi$, and $\xi$ project to the same section $\rho\in\Gamma(\pi_{X,\Sigma})$, that is,
\[
\pi_{\Sigma,L\left(\pi_{X,\Sigma}^*TX,\operatorname{Ad}P\right)}\circ\tilde\nabla^\mathcal{A}\xi=\pi_{\Sigma,\operatorname{Ad}P}\circ\tilde\nabla_X^\mathcal{A}\xi=\pi_{\Sigma,\operatorname{Ad}P}\circ\xi=\rho.
\]

Next, in the present situation, the covariant exterior derivative
\begin{equation}\label{d_A_special}d^ \mathcal{A} : = \mathbf{d}^{\tilde{\nabla}^{\cA}}: \Omega^1_{ \pi_{X, \Sigma}}(X, \operatorname{Ad}P) \rightarrow \Omega^2_{ \pi_{X, \Sigma}}(X, \operatorname{Ad}P)
\end{equation}
is attained from \eqref{tilde_covariant_nabla_exterior}.  That is,
\begin{equation}\label{d_A_special_formula}d^\cA \xi (u_x, v_x) = \tilde \nabla^{\cA}_{U} \lp \xi(V) \rp - \tilde \nabla^{\cA}_{V} \lp \xi(U) \rp - \xi\left(\lsb U, V \rsb \right),
\end{equation}
where $\xi \in  \Omega^1_{ \pi_{X, \Sigma}}(X, \operatorname{Ad}P)$, $u _x, v _x\in T _xX$, $U, V \in \mathfrak{X}(X)$ satisfying $U (x) = u _x$, $V (x) = v _x$.

\paragraph{Variations.} The wedge product \eqref{wedge_bundle_valued} in $\Omega^1(\Sigma,\operatorname{Ad}P)$ extends to elements $\alpha, \beta \in \Omega^1_{ \pi_{X, \Sigma}}(X,\operatorname{Ad}P)$ that project to the same element $\rho \in \Gamma( \pi_{X, \Sigma})$ by
\begin{equation}\label{wedge_X}(\alpha\wedge\beta)(u_x,v_x):=[\alpha(u_x),\beta(v_x)]_{\rho(x)} - [\alpha(v_x),\beta(u_x)]_{\rho(x)},
\end{equation}
where $u _x, v _x \in T _xX$.

\begin{proposition}\label{kinematic_second_component} 
Let $\sigma: X\to P$ be a smooth section of $\pi_{X,P}:P \to X$. Let $\cA$ be a connection on the principal bundle $\pi_{\Sigma, P}:P \to \Sigma$ and  $\bar\sig = \llsb \sig , \sig^*\cA\rrsb_{\mg} \in \Gamma\left(\pi_{X, L\lp \pi^*_{X,\Sig}TX, \operatorname{Ad}P\rp} \right)$ the reduced section. Then
\[
d^{\cA}\bar\sig + \half\bar\sig \wedge \bar \sig = \rho^*\tilde \cB,
\]
where $\rho$ is the section of $\pi_{X,\Sig}: \Sig \to X$ defined by $\rho:=\pi_{\Sig,P}\circ\sig$ and $\tilde \cB$ is the the $\operatorname{Ad}P$-valued two-form induced on $\Sigma$ by the curvature $\mathcal{B}=\mathbf{d}\mathcal{A}- \frac{1}{2}\lp \mathcal{A}\wedge\mathcal{A}\rp \in \Omega^2(P, \mathfrak{g})$.
\end{proposition}

\begin{proof}
For $U,V\in\mathfrak{X}(X)$, \eqref{covariant_der_big} gives
\begin{align*}
d^\cA\bar\sig(U,V) &= \tilde\nabla^\mathcal{A}_{U}(\bar\sig(V)) - \tilde\nabla^\mathcal{A}_{V}(\bar\sig(U)) - \bar\sig(\lsb U, V \rsb ),\\
&=\llsb\sigma,\mathbf{d}(\sigma^*\mathcal{A}(V))U-[\sigma^*\mathcal{A}(U),\sigma^*\mathcal{A}(V)]\rrsb_{\mg}\\
&\qquad\qquad
-\llsb\sigma,\mathbf{d}(\sigma^*\mathcal{A}(U))V-[\sigma^*\mathcal{A}(V),\sigma^*\mathcal{A}(U)]\rrsb_{\mg}\\
&\qquad\qquad
-\llsb\sigma,\sigma^*\mathcal{A}([U,V])\rrsb_{\mg}\\
&= \llsb \sig, \mathbf{d}(\sig^*\cA)(U,V)- 2\lsb\sig^*\cA(U),\sig^*\cA(V)\rsb\rrsb_{\mg},\\
&= \left[ \!\! \left[ \sig,  \sig^*\mathbf{d}\cA - \half\sig^*\left(\cA\wedge\cA\right)(U,V)\right] \!\! \right]_{\mg} -\left[ \!\! \left[ \sigma,\half\sig^*\left(\cA\wedge\cA\right)(U,V)\right] \!\! \right]_{\mg},\\
&= \llsb \sig, \sig^*\cB(U,V)\rrsb_{\mg} - \half(\bar\sig\wedge\bar\sig)(U,V),\\
&= \rho^*\tilde\cB(U,V) - \half(\bar\sig\wedge\bar\sig)(U,V).
\end{align*}
Thus we obtain the formula
\[
d^\cA \bar\sig + \half\bar\sig \wedge \bar\sig = \rho^*\tilde\cB
\]
as required. 
\end{proof}
\medskip

Note that $\bar\sigma\in\Omega^1(X,\rho^*\operatorname{Ad}P)$, where $\rho:=\pi_{\Sigma,P}\circ\sigma$ and $\rho^*\operatorname{Ad}P$ is the pull-back vector bundle over $X$. Therefore $d^\cA\bar\sigma, \bar\sigma\wedge\bar\sigma \in \Omega^2(X,\rho^*\operatorname{Ad}P)$. Since the reduced curvature $\tilde\cB$ belongs to the space $\Omega^1(\Sigma,\operatorname{Ad}P)$, the pullback $\rho^*\tilde\cB\in \Omega^2(X,\rho^*\operatorname{Ad}P)$. Thus, the formula $d^\cA \bar\sig + \bar\sig \wedge \bar\sig = \rho^*\tilde\cB$ is well-defined as an equality in $\Omega^2(X,\rho^*\operatorname{Ad}P)$.

\begin{corollary}\label{variation_second_component} 
Let $\sigma_\varepsilon$ be a smooth variation of the smooth section $\sigma:\bar{U}\rightarrow P$ and $\mathcal{A}$ a connection on the principal bundle $\pi_{\Sigma,P}:P\rightarrow\Sigma$. Then
\begin{align*}
\left.\frac{D^\mathcal{A}}{D\varepsilon}\right|_{\varepsilon=0}\llsb \sigma_\varepsilon(x),\sigma_\varepsilon^*\mathcal{A}\rrsb_{\mg}
&= \tilde\nabla^\mathcal{A}\llsb \sigma,\mathcal{A}(\delta\sigma)\rrsb_{\mg}+\llsb\sigma(x),\sigma^*(\mathbf{i}_{\delta\sigma}\mathcal{B}) (x)\rrsb_{\mg}\\
&\qquad -\llsb\sigma(x),[\mathcal{A}(\delta\sigma(x)),\sig^*\mathcal{A}(x)]\rrsb_{\mg},
\end{align*}
where $\mathcal{B}=\mathbf{d}\mathcal{A}-1/2\lp \mathcal{A}\wedge\mathcal{A}\rp$ is the curvature of the connection.

Thus, the infinitesimal variations of $\bar\sigma$ are of the form
\[
\delta^\mathcal{A}\bar\sigma=\tilde\nabla^\mathcal{A}\bar\eta-[\bar\eta,\bar\sigma]+\rho^*\left(\mathbf{i}_{\delta\rho}\tilde{\mathcal{B}}\right),
\]
where $\rho:=\pi_{\Sigma,L\left(\pi_{X,\Sigma}^*TX,\operatorname{Ad}P\right)}\circ\bar\sigma$, $\delta\rho$ is an arbitrary variation of $\rho$ vanishing on $\partial U$, $\bar\eta$ is an arbitrary section in $\Gamma(\pi_{X,\operatorname{Ad}P})$ that projects to $\rho$ and vanishes on $\partial U$, and $\tilde{\mathcal{B}}$ denotes the $\operatorname{Ad}P$-valued two-form induced on $\Sigma$ by the curvature $\mathcal{B}$.
\end{corollary}
\begin{proof}  The second formula is a direct consequence of the first. To prove the first, one could verify the identity in local bundle charts. We prefer a global
proof based on the previous lemma.

Extending the bundle geometry in order to explicitly take account of variations achieves the objective.  Consider $\hat P = \mR \times P$ and $\hat X = \mR \times X$ with the projection $\hat \pi_{\hat X, \hat P}(\varepsilon, p) = (\varepsilon, \pi_{X,P}(p))$.  Smooth sections of $\hat\pi_{\hat X, \hat P}$, $\hat \sig: \hat X \to \hat P$ are in bijective correspondence with smooth variations of smooth sections of $\pi_{X,P}$, as follows:
\[
\hat \sig(\varepsilon, x) = (\varepsilon, \sig_ \varepsilon(x)).
\]
Let $G $ act on $\hat{P} $ by extending the action trivially to the $\mathbb{R}$-factor. Thus, $\hat{ \Sigma} : \hat{P}/G = \mathbb{R} \times \Sigma$ and $\operatorname{Ad}\hat{P} = \mathbb{R} \times \operatorname{Ad} P $.
Since $\hat{\rho} : = \pi_{\hat{X}, \hat{ \Sigma}} \circ \hat{\sigma}$ it is clear that $\hat{\rho}( \varepsilon, x ) = ( \varepsilon , \rho_\varepsilon(x))$.  Similarly, the connection $\mathcal{A}$ extends to a connection $\hat{ \mathcal{A}} \in \Omega^1(\hat{P}, \mathfrak{g}) $ by setting $\hat{ \mathcal{A}}( \partial_ \varepsilon, u _p): = \mathcal{A}(u_p) $, for any $u_p \in T_pP$.

The section $\hat{\sigma}:\hat{X} \rightarrow \hat{P}$ induces the reduced section $\bar{\hat{\sigma}}$ (see \eqref{def_sigma_bar} for the general definition) whose explicit expression may be computed as follows: For $( \varepsilon,x) \in \hat{X} $, $u _x\in T _xX $, letting $\delta \sigma_ \varepsilon : = \frac{d}{d\varepsilon} \sigma_ \varepsilon$, and using
\[
T\hat{ \sigma}(\partial_\varepsilon, u_x) = ( \partial_\varepsilon, T \sigma _\varepsilon (u_x) + \delta \sigma_\varepsilon(x)),
\]
generates 
\begin{align}
\label{sigma_hat_bar_formula}
\bar{\hat{\sigma}}( \partial_ \varepsilon, u _x) : &= \left [ \!\! \left [ \hat{ \sigma}( \varepsilon, x), \hat {\mathcal{A}} \left(T\hat{ \sigma}(\partial_\varepsilon, u_x) \right)\right ] \!\! \right ]_{\mg}
= \llsb( \varepsilon, \sigma_ \varepsilon(x)), \mathcal{A}\left(T \sigma_ \varepsilon(u_x) + \delta \sigma_ \varepsilon(x) \right) \rrsb_{\mg} \nonumber \\
& = \left( \varepsilon, \bar{\sigma}_ \varepsilon(u_x) + \llsb \sigma_ \varepsilon(x), \mathcal{A}(\delta\sigma_ \varepsilon(x))\rrsb_{\mg} \right).
\end{align}

The required formula is attained by evaluating the identity in Proposition \ref{kinematic_second_component},
\[
d^{\hat{ \mathcal{A}}} \bar{\hat{\sigma}} + \frac{1}{2}\bar{\hat{\sigma}} \wedge \bar{\hat{\sigma}} = \hat{\rho} ^\ast \tilde{\hat{\mathcal{B}}},
\]
on the pair of vectors $(\partial_\varepsilon, 0)$, $(\partial_\varepsilon, v_x)$ for $v_x \in T_xX$.  A direct computation shows that
\begin{align*}
\hat{\rho} ^\ast \tilde{\hat{\mathcal{B}}} \left((\partial_\varepsilon, 0), (\partial_\varepsilon, v_x) \right) 
& = \left(\varepsilon, \llsb \sigma_ \varepsilon(x), \mathcal{B} \left( \delta\sigma_ \varepsilon(x), T \sigma_ \varepsilon(v_x) \right) \rrsb_{\mg} \right),\\
\frac{1}{2}\left(\bar{\hat{\sigma}} \wedge \bar{\hat{\sigma}}\right) \left((\partial_\varepsilon, 0), (\partial_\varepsilon, v_x) \right) 
& = \left(\varepsilon, \llsb \sigma_ \varepsilon(x), \left[\mathcal{A}(\delta\sigma_ \varepsilon(x)), \mathcal{A}\left(T \sigma_ \varepsilon(v_x) \right) \right] \rrsb_{\mg} \right).
\end{align*}
To calculate $d^{\hat{\mathcal{A}}} \bar{\hat{\sigma}}\left((\partial_\varepsilon, 0), (\partial_\varepsilon, v_x) \right)$, let $V \in \mathfrak{X}(X) $ be such that  $V(x) = v_x$ and use \eqref{d_A_special_formula}, \eqref{sigma_hat_bar_formula}, and $\left[\left((\partial_\varepsilon, 0), (\partial_\varepsilon, V) \right)  \right] = 0 $ to get
\begin{align*}
d^{\hat{ \mathcal{A}}} \bar{\hat{\sigma}}\left((\partial_\varepsilon, 0), (\partial_\varepsilon, v_x) \right) 
& = \left(\tilde{\nabla}^{\hat{ \mathcal{A}}}_{(\partial_\varepsilon, 0)} \left(\bar{\hat{ \sigma}}(\partial_\varepsilon, V) \right) \right) ( \varepsilon, x)
- \lp \tilde{\nabla}^{\hat{ \mathcal{A}}}_{(\partial_\varepsilon, v_x)} \left(\bar{\hat{ \sigma}}(\partial_\varepsilon, 0) \right)\rp ( \varepsilon, x) \\
& \qquad - \bar{\hat{\sigma}} \left[\left((\partial_\varepsilon, 0), (\partial_\varepsilon, V) \right)  \right] (\varepsilon, x) \\
& = \tilde{\nabla}^{\hat{ \mathcal{A}}}_{(\partial_\varepsilon, 0)} (\varepsilon , \bar{ \sigma}_ \varepsilon(V)) ( \varepsilon, x) - 
 \tilde{\nabla}^{\hat{ \mathcal{A}}}_{(0, v_x)} \left(\varepsilon, \llsb \sigma_\varepsilon(x), \mathcal{A} (\delta\sigma_\varepsilon(x))\rrsb_{\mg}  \right) \\
&= \frac{D^{\hat{\mathcal{A}}}}{ D\varepsilon} ( \varepsilon, \bar{ \sigma} _\varepsilon(v_x) ) - \left(\varepsilon, \tilde{ \nabla}^{ \mathcal{A}}_{v_x} \llsb \sigma_\varepsilon(x), \mathcal{A} (\delta\sigma_\varepsilon(x))\rrsb_{\mg} \right)\\
& = \left(\varepsilon, \frac{ D^{ \mathcal{A}}}{ D\varepsilon} \bar{ \sigma} _\varepsilon(v_x) - \tilde{ \nabla}^{ \mathcal{A}}_{v_x} \llsb \sigma_\varepsilon(x), \mathcal{A} (\delta\sigma_\varepsilon(x))\rrsb_{\mg} \right).
\end{align*}
The last three identities prove the first stated formula. 
\end{proof}

\medskip

A covariant derivative $\nabla^\Sigma$ on the tangent bundle $\tau_\Sigma:T\Sigma\rightarrow\Sigma$ is needed in order to compute the variation of $T_x\rho$. Given $\nabla^\Sigma$, \eqref{nabla_tilde_general_def} defines the $\nabla^\Sigma$-derivative $\tilde{\nabla}^\Sigma$ which acts on functions $X \rightarrow T \Sigma$ and thus \eqref{tilde_covariant_nabla_exterior} provides the operator 
\[
d^\Sigma : = \mathbf{d}^{\tilde{\nabla}^ \Sigma}: \Omega^1_{\pi_{X, \Sigma}}(X, T \Sigma) \rightarrow \Omega^2_{\pi_{X, \Sigma}}(X, T \Sigma)
\]
defined by
\begin{equation}\label{def_d_Sigma}d^\Sigma\lambda(U,V) = \tilde\nabla^{\Sig}_{U}(\lambda(V)) - \tilde\nabla^{\Sig}_{V}(\lambda(U)) - \lambda(\lsb U,V\rsb),
\end{equation}
where $\lambda \in \Omega^1_{\pi_{X, \Sigma}}(X, T \Sigma)$ and $U, V \in \mathfrak{X}(X)$. In particular, since $J^1\Sigma\subset L(\pi_{X,\Sigma}^*TX,T\Sigma)$, sections of $J ^1 \Sigma \rightarrow X $ are necessarily sections of $L(\pi_{X,\Sigma}^*TX,T\Sigma) \rightarrow X$, that is, elements of $\Omega^1_{\pi_{X, \Sigma}}(X, T \Sigma)$ and thus $d^ \Sigma$ operates on sections of the bundle $J ^1 \Sigma \rightarrow X $. 

This differential operator satisfies the following property.
\begin{proposition}\label{kinematic_first_component}  
Let $\rho:X \to \Sigma$ be a smooth section of $\pi_{X, \Sigma}$.  Then
\begin{align*}
d^\Sigma (j^1\rho) = \rho^*T^\Sigma.
\end{align*}
where $T^\Sigma(U,V)=\nabla^\Sigma_UV-\nabla^\Sigma_VU-[U,V]$ is the \textit{torsion tensor} of the connection $\nabla^\Sigma$.
\end{proposition}
\begin{proof} Recall that the section $j^1\rho: X \rightarrow J^1\Sigma$ is interpreted in this formula in the following way. Given $s \in \Sigma$, let $x: = \pi_{X, \Sigma}(s) \in X $ and so $j ^1 \rho(x) = T_x \rho: T_x X \rightarrow T_{ \rho(x)} \Sigma$, that is, one thinks of $j^1 \rho$ as an element of $\Omega^1_{ \pi_{X, \Sigma}}(X, T \Sigma)$. Given $u_x, v_x\in T_xX$ and having chosen two vector fields $U, V \in \mathfrak{X}(X) $ such that $U(x)=u_x$ and $V(x)=u_x$, \eqref{def_d_Sigma} and \eqref{nabla_tilde_general_def} confer
\begin{align*}
d^\Sigma (j^1\rho)(u_x,v_x) &= \tilde\nabla^\Sig_{U} (T\rho\lp V\rp) - \tilde\nabla^\Sig_{V} (T\rho\lp U\rp) - T\rho\lp \lsb U,V\rsb \rp\\
&=\left.\frac{D^{\nabla^\Sigma}}{Dt}\right|_{t=0}T\rho(V(c_1(t)))-\left.\frac{D^{\nabla^\Sigma}}{Dt}\right|_{t=0}T\rho(U(c_2(t)))-T\rho\lp \lsb U,V\rsb \rp,
\end{align*}
where $c_1(t) $ and $c_2(t) $ are curves in in $X $ such that $c_1(0)= c_2(0) = x $ and $\dot{c}_1(0) = u_x$, $\dot{c}_2(0) = v_x$.
Since $\rho$ is a section of $\pi_{X,\Sigma}$, it is an embedding and the image $\rho(X)$ is a submanifold of $\Sigma$. Thus, there exists vector fields $\bar U, \bar V\in\mathfrak{X}(\Sigma)$ such that $\bar U(\rho(x))=T\rho(U(x))$ and $\bar V(\rho(x))=T\rho(V(x))$. Accordingly,
\begin{align*}
\rho^*T^\Sigma(u_x,v_x)&=T^\Sigma(T\rho(U(x)),T\rho(V(x)))=T^\Sigma(\bar U,\bar V)(\rho(x))\\
&=\left(\nabla^\Sigma_{\bar U}\bar V-\nabla^\Sigma_{\bar V}\bar U-[\bar U,\bar V]\right)(\rho(x))\\
&=\left.\frac{D^{\nabla^\Sigma}}{Dt}\right|_{t=0}\bar V(\rho(c_1(t)))-\left.\frac{D^{\nabla^\Sigma}}{Dt}\right|_{t=0}\bar U(\rho(c_2(t)))-[\bar U,\bar V](\rho(x))\\
&=\left.\frac{D^{\nabla^\Sigma}}{Dt}\right|_{t=0}T\rho(V(c_1(t)))-\left.\frac{D^{\nabla^\Sigma}}{Dt}\right|_{t=0}T\rho(U(c_2(t)))-T\rho\lp \lsb U,V\rsb \rp,
\end{align*}
which proves the statement.
\end{proof}
\medskip

The next result may be obtained using the previous formula by extending the bundle geometry to $\hat{P}$ as was done in the proof of Corollary \ref{variation_second_component} using Proposition \ref{kinematic_second_component}. This time, however, we provide a different proof based on a standard formula for the torsion.

\begin{corollary}\label{variation_first_component} Let $\rho_\varepsilon$ be a smooth variation of the section $\rho:\bar{U}\rightarrow \Sigma$, $u_x\in T_xX$, and let  $\nabla^\Sigma$ be a covariant derivative on $T\Sigma$. Then
\[
\left.\frac{D^{\nabla^\Sigma}}{D\varepsilon}\right|_{\varepsilon=0}T_x\rho_\varepsilon(u_x)=\tilde\nabla^\Sigma_{u_x}\delta\rho+T^\Sigma\left(\delta\rho(x),T_x\rho(u_x)\right).
\]
where $\tilde\nabla^\Sigma$ is the $\nabla^\Sigma$-derivative and $T^\Sigma(U,V)=\nabla^\Sigma_UV-\nabla^\Sigma_VU-[U,V]$ is the \textit{torsion tensor} of the connection $\nabla^\Sigma$.
\end{corollary}
\begin{proof} This is a direct consequence of the formula
\[
\frac{D^{\nabla^\Sigma}}{D\varepsilon}\frac{d}{dt}\alpha(\varepsilon,t)-\frac{D^{\nabla^\Sigma}}{D t}\frac{d}{d\varepsilon}\alpha(\varepsilon,t)=T^\Sigma\left(\frac{d}{d\varepsilon}\alpha(\varepsilon,t),\frac{d}{d t}\alpha(\varepsilon,t)\right),
\]
where $(\varepsilon,t)\in I\times J\mapsto \alpha(\varepsilon,t)\in \Sigma$ is a smooth smooth function. Here it suffices to choose $\alpha(\varepsilon,t)=\rho_\varepsilon(c(t))$, where $c$ is a smooth curve in $X$ such that $\dot c(0)=u_x$.
\end{proof}

\medskip

For simplicity, we will always choose a torsion free connection $\nabla^\Sigma$. In this case, the previous formulas simplify to
\[
d^\Sigma (j^1\rho)=0\quad\text{and}\quad \left.\frac{D^{\nabla^\Sigma}}{D\varepsilon}\right|_{\varepsilon=0}T_x\rho_\varepsilon(u_x)=\tilde\nabla^\Sigma_{u_x}\delta\rho.
\]

\subsection{The Lagrange-Poincar\'e field equations}

Let $l:J^1\Sigma\oplus_\Sigma L\left(\pi_{X,\Sigma}^*TX,\operatorname{Ad}P\right)\rightarrow\mathbb{R}$ be the reduced Lagrangian (see \eqref{three_Lagrangians}). This Section computes the Lagrange-Poincar\'e equations given by the variational principle
\[
0=\left.\frac{d}{d\varepsilon}\right|_{\varepsilon=0}\int_U\mathcal{L}(j^1\sigma_\varepsilon)
=\left.\frac{d}{d\varepsilon}\right|_{\varepsilon=0}\int_Ul(j^1\rho_ \varepsilon,\bar\sigma_\varepsilon)\mu.
\]
An affine connection on the affine bundle $J^1\Sigma\oplus_\Sigma L\left(\pi_{X,\Sigma}^*TX,\operatorname{Ad}P\right)\rightarrow\Sigma$ is required in order to obtain explicit formulas.  Since the principal connection $\cA$ brings a covariant derivative on $\operatorname{Ad}P\rightarrow\Sigma$, it suffices to choose a covariant derivative $\nabla^\Sigma$ on the vector bundle $T\Sigma\rightarrow \Sigma$. This induces a connection on $J^1\Sigma$ given by
\begin{equation}\label{connection_j_one_sigma}\nabla^{J^1\Sigma}_Z\gamma:=\operatorname{ver}^{\gamma}\circ \nabla^L_Z\gamma,
\end{equation}
where $\operatorname{ver}^\gamma$ is the vertical projection associated to the section $\gamma\in\Gamma(\pi_{\Sigma,J^1\Sigma})$, interpreted as a connection on $\pi_{X,\Sigma}$, and $Z$ is a vector field on $\Sigma$. Here $\nabla^L$ denotes the covariant derivative induced on $L\left(\pi^*_{X,\Sigma}TX,T\Sigma\right)$, from $\nabla^\Sigma$ and from a covariant derivative $\nabla^X$ on $TX$, that is,
\begin{equation}\label{L-derivative}
\left(\nabla ^L_\xi \ell \right)(U)(s) : = \nabla _ \xi^ \Sigma (\ell (U)) (s) 
- \ell \left(\nabla _{T \pi_{X, \Sigma} ( \xi(s))} U \right),
\end{equation}
where $\ell$ is a section of the vector bundle $L\left(\pi^*_{X,\Sigma}TX,T\Sigma\right)\rightarrow  \Sigma$, $\xi \in \mathfrak{X} ( \Sigma) $, $U \in \mathfrak{X}(X) $, $s \in \Sigma$, and recall that $\ell (U) (s) : = \ell(s) (U( \pi_{X, \Sigma} (s)))$.
However, the final result only depends on $\nabla^\Sigma$ and not on $\nabla^X$, see \cite{JaMo1996}. In this paper it is also shown that if $\nabla^\Sigma$ is projectable onto a covariant derivative on $X$, then $\nabla^{J^1\Sigma}_Z\gamma$ is an affine connection.

Thus assuming a projectable covariant derivative $\nabla^\Sigma$ is given, an affine connection on $J^1\Sigma\oplus_\Sigma L\left(\pi_{X,\Sigma}^*TX,\operatorname{Ad}P\right)$ is obtained. Given a smooth function $l$ on this affine bundle, define the fiber derivatives
\[
\left\langle\frac{\delta l}{\delta j^1\rho}(x),v\right\rangle:=\left.\frac{d}{d\varepsilon}\right|_{\varepsilon=0}l(j^1\rho(x)+\varepsilon v,\bar\sigma(x)),\quad \left\langle\frac{\delta l}{\delta \bar\sigma}(x),w\right\rangle:=\left.\frac{d}{d\varepsilon}\right|_{\varepsilon=0}l(j^1\rho(x),\bar\sigma(x)+\varepsilon w),
\]
where $v \in L(T_xX,V_{\rho(x)}\Sigma)$ and $w \in L(T_xX,\operatorname{Ad}P_{\rho(x)})$ are arbitrary vectors. Note that $\delta l/\delta j^1\rho$ and $\delta l/\delta \bar\sigma$ are sections of the bundles $L\left(\pi_{X,\Sigma}^*T^*X,V\Sigma^*\right)\rightarrow X$ and $L\left(\pi_{X,\Sigma}^*T^*X,\operatorname{Ad}P^*\right)\rightarrow X$; both project to $\rho$. The derivative with respect to $\rho$ is the horizontal derivative defined at $(j^1\rho(x),\bar\sigma(x))$ by
\begin{equation}\label{hor_deriv}
\left\langle\frac{\delta l}{\delta\rho}(x),u\right\rangle:=\left.\frac{d}{d\varepsilon}\right|_{\varepsilon=0}l\left(\sigma(\varepsilon)^h_{(j^1\rho(x),\bar\sigma(x))}\right),
\end{equation}
where $\sigma(\varepsilon)$ is a curve in $\Sigma$ such that $\dot\sigma(0)=u\in T_{\rho(x)}\Sigma$, and $\sigma(\varepsilon)^h_{(j^1\rho(x),\bar\sigma(x))}$ is the unique horizontal curve starting at $(j^1\rho(x),\bar\sigma(x))$ and projecting to $\sigma(\varepsilon)$.

Consider a variation $\sigma_\varepsilon$ of a given local section $\sigma:\bar{U}\rightarrow P$ and the reduced section $\bar\sigma_\varepsilon$. Employing the decomposition of the $\varepsilon$-derivative into its vertical and horizontal parts yields
\begin{align}\label{variations}
&\left.\frac{d}{d\varepsilon}\right|_{\varepsilon=0}\int_Ul(j^1\rho_\varepsilon(x),\bar\sigma_\varepsilon(x))\mu
=\int_U\mathbf{d}l\left(\left.\frac{d}{d\varepsilon}\right|_{\varepsilon=0}
\left(j^1 \rho_\varepsilon(x),\bar\sigma_\varepsilon(x)\right)\right)\mu
\nonumber\\
&=\int_U\left\langle\frac{\delta l}{\delta\rho}(x),\left.\frac{d}{d\varepsilon}\right|_{\varepsilon=0}\rho_\varepsilon(x)\right\rangle\mu+
\int_U\left\langle\frac{\delta l}{\delta j^1\rho}(x),\left.\frac{D^{J^1\Sigma}}{D\varepsilon}\right|_{\varepsilon=0}j^1\rho_\varepsilon(x)\right\rangle\mu\\
&\qquad+\int_U\left\langle\frac{\delta l}{\delta\rho}(x), \left.\frac{D^L}{D\varepsilon}\right|_{\varepsilon=0}\bar\sigma_\varepsilon(x)  \right\rangle\mu\nonumber,
\end{align}
where $D^{J^1\Sigma}/D\varepsilon$ and $D^L/D\varepsilon$ denote the covariant derivatives associated to the connection $\nabla^{J^1\Sigma}$ on $J^1\Sigma\rightarrow\Sigma$ and to the induced covariant derivative on $L\left(\pi_{X,\Sigma}^*TX,\operatorname{Ad}P\right)\rightarrow\Sigma$, respectively.

The second term may be computed using the following relation:
\begin{equation}\label{expression1}
\left(\left.\frac{D^{J^1\Sigma}}{D\varepsilon}\right|_{\varepsilon=0}j^1\rho_\varepsilon(x)\right)(v_x)=\left.\frac{D^{T\Sigma}}{D\varepsilon}\right|_{\varepsilon=0}\left(j^1\rho_\varepsilon(x)(v_x)\right).
\end{equation}
This relation is obtained from the definition of the induced covariant derivative $D^L/D\varepsilon$ on  $L(\pi_{X,\Sigma}^*TX,T\Sigma)$.  Given a curve $\gamma_\varepsilon\in J^1\Sigma$, \eqref{L-derivative} shows that
\[
\left(\left.\frac{D^L}{D\varepsilon}\right|_{\varepsilon=0}\gamma_\varepsilon\right)(v_x)=\left(\left.\frac{D^{T\Sigma}}{D\varepsilon}\right|_{\varepsilon=0}(\gamma_\varepsilon\!\cdot\!v_\varepsilon)\right)-\gamma_0\!\cdot\!\left(\left.\frac{D^{TX}}{D\varepsilon}\right|_{\varepsilon=0}v_\varepsilon\right),
\]
where $v_\varepsilon\in TX$ is a curve such that $v_\varepsilon\in T_{x_\varepsilon}X$ and $x_\varepsilon\in X$ is such that $\gamma_\varepsilon\in L(T_{x_\varepsilon}X,T_{s_\varepsilon}\Sigma)$.  In the present case $\gamma_\varepsilon=j^1\rho_\varepsilon(x)$ and variations in $TX$ are not considered, so $x_\varepsilon=x$ and $v_\varepsilon=v_x$. Thus
\begin{equation}\label{intermediate_step}
\left(\left.\frac{D^L}{D\varepsilon}\right|_{\varepsilon=0}(j^1\rho_\varepsilon(x))\right)(v_x)=\left.\frac{D^{T\Sigma}}{D\varepsilon}\right|_{\varepsilon=0}(j^1\rho_\varepsilon(x)\!\cdot\!v_x).
\end{equation}
Denoting the connector map of $\nabla^\Sigma$ by $K^{T\Sigma}$ and recalling that $\nabla^\Sigma$ is projectable allows the following calculation:
\begin{align*}
&T\pi_{X, \Sigma}\left(\left.\frac{D^{T\Sigma}}{D\varepsilon}\right|_{\varepsilon=0}(j^1\rho_\varepsilon(x)\!\cdot\!v_x)\right)=T\pi_{X,\Sigma}\left(K^{T\Sigma}\left(\left.\frac{d}{d\varepsilon}\right|_{\varepsilon=0}(j^1\rho_\varepsilon(x)\!\cdot\!v_x)\right)\right)\\
&=K^{TX}\left(TT\pi_{X,\Sigma}\left(\left.\frac{d}{d\varepsilon}\right|_{\varepsilon=0}(j^1\rho_\varepsilon(x)\!\cdot\!v_x)\right)\right)=K^{TX}\left(\left.\frac{d}{d\varepsilon}\right|_{\varepsilon=0}T\pi_{X,\Sigma}\left(j^1\rho_\varepsilon(x)\!\cdot\!v_x\right)\right)\\
&=K^{TX}\left(\left.\frac{d}{d\varepsilon}\right|_{\varepsilon=0}v_x\right)=0.
\end{align*}
This proves that the expression \eqref{intermediate_step} is vertical. Thus, by the definition \eqref{connection_j_one_sigma} of $\nabla^{J^1\Sigma}$, the identity \eqref{expression1} is proved.

The third term in equation \eqref{variations} may be evaluated using the equality
\begin{equation}\label{expression2}
\left(\left.\frac{D^L}{D\varepsilon}\right|_{\varepsilon=0}\bar\sigma_\varepsilon(x)\right)(v_x)=\left.\frac{D^\mathcal{A}}{D\varepsilon}\right|_{\varepsilon=0}\bar\sigma_\varepsilon(x)(v_x),
\end{equation}
whose proof is similar to that of \eqref{expression1}.

Using \eqref{expression1}, \eqref{expression2}, and Lemmas \ref{variation_second_component}, \ref{variation_first_component}, the expression \eqref{variations} may be rewritten
\begin{align*}
&\left.\frac{d}{d\varepsilon}\right|_{\varepsilon=0}\int_Ul(j^1\rho_\varepsilon(x),\bar\sigma_\varepsilon(x))\mu\\
&=\int_U\left\langle\frac{\delta l}{\delta\rho},\delta\rho\right\rangle\mu+
\int_U\left\langle\frac{\delta l}{\delta j^1\rho},\tilde\nabla^\Sigma\delta\rho+T^\Sigma\left(\delta\rho,T\rho\right)\right\rangle\mu\\
&\qquad+\int_U\left\langle\frac{\delta l}{\delta\bar\sigma}, \tilde\nabla^\mathcal{A}\bar\eta-[\bar\eta,\bar\sigma]+\rho^*\left(\mathbf{i}_{\delta\rho}\tilde{\mathcal{B}}\right)\right\rangle\mu\\
&=\int_U\left\langle -\operatorname{div}^\mathcal{A}\frac{\delta l}{\delta\bar\sigma}+\operatorname{ad}^*_{\bar\sigma}\frac{\delta l}{\delta\bar\sigma},\bar\eta\right\rangle\mu\\
&\qquad+\int_U\left\langle\frac{\delta l}{\delta\rho}-\operatorname{div}^\Sigma\frac{\delta l}{\delta j^1\rho}-\left\langle\frac{\delta l}{\delta\bar\sigma}, \mathbf{i}_{T\rho}\tilde{\mathcal{B}}\right\rangle+\left\langle\frac{\delta l}{\delta j^1\rho}, \mathbf{i}_{T\rho}T^\Sigma\right\rangle,\delta\rho\right\rangle\mu
\end{align*}

Since $\delta\rho$ and $\bar\eta$ are arbitrary, this results in the \textit{vertical} and \textit{horizontal Lagrange-Poincar\'e equations} given by
\begin{equation}\label{LP_eqn_torsion}
\operatorname{div}^\mathcal{A}\frac{\delta l}{\delta\bar\sigma}-\operatorname{ad}^*_{\bar\sigma}\frac{\delta l}{\delta\bar\sigma}=0\quad\text{and}\quad \frac{\delta l}{\delta\rho}-\operatorname{div}^\Sigma\frac{\delta l}{\delta j^1\rho}+\left\langle\frac{\delta l}{\delta j^1\rho}, \mathbf{i}_{T\rho}T^\Sigma\right\rangle=\left\langle\frac{\delta l}{\delta\bar\sigma}, \mathbf{i}_{T\rho}\tilde{\mathcal{B}}\right\rangle,
\end{equation}
respectively, where the second equation has to be considered as an equation in $V_{\rho(x)}\Sigma^*$. For simplicity, we will suppose that $\nabla^\Sigma$ is torsion free.

Here $\operatorname{div}^\mathcal{A}$ denotes the divergence associated with $\tilde\nabla^\mathcal{A}$,
\[
\operatorname{div}^\mathcal{A}:\Gamma\left(\pi_{X,L\left(\pi_{X,\Sigma}^*T^*X,\operatorname{Ad}P^*\right)}\right)\rightarrow \Gamma\left(\pi_{X,\operatorname{Ad}P^*}\right),
\]
which is defined as minus the adjoint differential operator to $\tilde\nabla^\mathcal{A}$:
\[
\int_X\left\langle\operatorname{div}^\mathcal{A}l(x),\xi(x)\right\rangle\mu=-\int_X\left\langle l(x),\tilde\nabla^\mathcal{A}\xi(x)\right\rangle\mu,
\]
for all $l\in\Gamma\left(\pi_{X,L\left(\pi_{X,\Sigma}^*T^*X,\operatorname{Ad}P^*\right)}\right)$ and $\xi\in\Gamma\left(\pi_{X,\operatorname{Ad}P}\right)$ such that $\pi_{\Sigma,L\left(\pi_{X,\Sigma}^*T^*X,\operatorname{Ad}P^*\right)}\circ l=\pi_{\Sigma,\operatorname{Ad}P}\circ\xi.$
In the vertical equation, $\operatorname{ad}^*$ denotes the map
\[
\operatorname{ad}^*:\Gamma\left(\pi_{X,L\left(\pi_{X,\Sigma}^*TX,\operatorname{Ad}P\right)}\right)\times\Gamma\left(\pi_{X,L\left(\pi_{X,\Sigma}^*T^*X,\operatorname{Ad}P^*\right)}\right)\rightarrow \Gamma\left(\pi_{X,\operatorname{Ad}P^*}\right),\quad(\bar\sigma,\bar\mu)\mapsto \operatorname{ad}^*_{\bar\sigma}\bar\mu,
\]
well-defined when $\pi_{\Sigma,L\left(\pi_{X,\Sigma}^*T^*X,\operatorname{Ad}P^*\right)}\circ\bar\mu=\pi_{\Sigma,L\left(\pi_{X,\Sigma}^*TX,\operatorname{Ad}P\right)}\circ\bar\sigma$. Similarly, the operator
\begin{equation}\label{div_sigma}
\operatorname{div}^\Sigma:\Gamma\left(\pi_{X,L\left(\pi_{X,\Sigma}^*T^*X,V\Sigma^*\right)}\right)\rightarrow \Gamma\left(\pi_{X,V\Sigma^*}\right)
\end{equation}
is the divergence associated to the $\nabla^\Sigma$-derivative $\tilde\nabla^\Sigma$ restricted to vertical valued sections:
\[
\tilde\nabla^\Sigma:\Gamma(\pi_{X,V\Sigma})\rightarrow \Gamma\left(\pi_{X,L\left(\pi_{X,\Sigma}^*TX,V\Sigma\right)}\right).
\]
Note that such a restriction is possible since $\nabla^\Sigma$ is projectable.
The results obtained above are summarized in the following theorem.

\begin{theorem}\label{EP_field_reduction} Let $\pi_{X,P}:P\rightarrow X$ be a locally trivial fiber bundle over an oriented manifold with volume form $\mu$. Let $L:J^1P\rightarrow \mathbb{R}$ be a Lagrangian which is invariant under a free and proper left action $\Phi:G\times P\rightarrow P$ such that
\[
\pi_{X,P}\circ\Phi_g=\pi_{X,P},\quad\text{for all $g\in G$}.
\]
Let $\pi_{\Sigma,P}:P\rightarrow\Sigma:=P/G$ be the associated principal bundle.

Fix a connection $\mathcal{A}$ on $\pi_{\Sigma,P}$ and let $l:J^1\Sigma\oplus_\Sigma L\left(\pi_{X,\Sigma}^*TX,\operatorname{Ad}P\right)\rightarrow\mathbb{R}$ be the reduced Lagrangian induced on the quotient by means of the identification \eqref{definition_beta}. Let $\sigma:\bar{U}\rightarrow P$ be a smooth local section of $\pi_{X,P}$, define the reduced local section $\bar\sigma$ of $\pi_{X,L\left(\pi_{X,\Sigma}^*TX,\operatorname{Ad}P\right)}$ by
\[
\bar\sigma(x)=\llsb \sigma(x),\mathcal{A}(T_x\sigma(\_\,)\rrsb_{\mg},
\]
and the local section $\rho:=\pi_{\Sigma,P}\circ\sigma$ of $\pi_{X,\Sigma}$. Fix a projectable covariant derivative $\nabla^\Sigma$ on $T\Sigma$ and suppose, for simplicity, that $\nabla^\Sigma$ is torsion free. Then the following are equivalent:
\begin{itemize}
\item[{\bf i}] The variational principle
\[
\delta\int_UL(j^1\sigma)\mu=0,
\]
holds for arbitrary vertical variations $\delta\sigma$ vanishing on $\partial U$.
\item[{\bf ii}] The section $\sigma$ satisfies the covariant Euler-Lagrange equations for $L\mu$.
\item[{\bf ii}] The variational principle
\[
\delta\int_Xl(j^1\rho,\bar\sigma)\mu=0,
\]
holds, for variations of the form $\delta^\mathcal{A}\bar\sigma=\nabla^\mathcal{A}\bar\eta-[\bar\eta,\bar\sigma]+\tilde{\mathcal{B}}(\delta\rho,T\rho)$, where $\delta\rho$ is an arbitrary variation of $\rho$ vanishing on $\partial U$ and $\bar\eta$ is an arbitrary section of $\pi_{X,\operatorname{Ad}P}$ vanishing on $\partial U$ and such that $\pi_{\Sigma,\operatorname{Ad}P}\circ\bar\eta=\rho$.
\item[{\bf iv}] The section $\bar\sigma$ satisfies the Lagrange-Poincar\'e field equations
\begin{equation}\label{LP_field_eqn}
\left\{\begin{array}{ll}
\vspace{0.2cm}\displaystyle\frac{\delta l}{\delta\rho}-\operatorname{div}^\Sigma\frac{\delta l}{\delta j^1\rho}=\left\langle\frac{\delta l}{\delta\bar\sigma}, \mathbf{i}_{T\rho}\tilde{\mathcal{B}}\right\rangle,\\
\displaystyle\operatorname{div}^\mathcal{A}\frac{\delta l}{\delta\bar\sigma}-\operatorname{ad}^*_{\bar\sigma}\frac{\delta l}{\delta\bar\sigma}=0.
\end{array}\right.
\end{equation}
\end{itemize}
\end{theorem}

In the case of a connection $\nabla^\Sigma$ with torsion, a term involving the torsion tensor has to be added in the horizontal Lagrange-Poincar\'e field equations, see \eqref{LP_eqn_torsion}.

\medskip

The Lagrange-Poincar\'e field equations are now examined in the particular cases mentioned before.
\begin{itemize}
\item[{\bf i}] If $G=\{1\}$ then $\Sigma=P$, $\rho=\sigma$, $l=L$, and there is no reduction. In this case \eqref{LP_field_eqn} becomes
\begin{equation}\label{eq_case_iii}\frac{\delta L}{\delta\sigma}-\operatorname{div}^P\frac{\delta L}{\delta j^1\sigma}=0,
\end{equation}
which is just a restatement of the covariant Euler-Lagrange equations, using a projectable and torsion free covariant derivative $\nabla^P$ on $TP\rightarrow P$.
\item[{\bf ii}] If $\pi_{X,P}:P \rightarrow X $ is a principal bundle and the symmetry group is the structure group, then $\Sigma=X$ and the section $\rho$ is absent since it is the identity on $X$. Therefore, the reduced variation reads $\delta^\mathcal{A}\bar\sigma=\nabla^\mathcal{A}\bar\eta-[\bar\eta,\bar\sigma]$, where $\bar\eta$ is an arbitrary section of $\pi_{X,\operatorname{Ad}P}$, and the Lagrange-Poincar\'e field equations \eqref{LP_field_eqn} read
\[
\operatorname{div}^\mathcal{A}\frac{\delta l}{\delta\bar\sigma}-\operatorname{ad}^*_{\bar\sigma}\frac{\delta l}{\delta\bar\sigma}=0.
\]
Thus the covariant Euler-Poincar\'e equations are recovered; see Theorem 3.1 of \cite{CaRaSh2000}.
\item[{\bf iii}] If $\pi_{X,P}:P\rightarrow X$ is a principal bundle whose structure group contains the symmetry group as a subgroup, the equations \eqref{LP_field_eqn} coincide with the equations (4.11) obtained in \cite{CaRa2003}.
\item[{\bf iv}] If $P=\mathbb{R}\times Q$ where $\pi_{M,Q}:Q\rightarrow M$ is a $G$-principal bundle  then $\Sigma=\mathbb{R}\times M$. The sections $\sigma\in\Gamma(\pi_{\mathbb{R},P})$ and $\rho\in\Gamma(\pi_{\mathbb{R},\Sigma})$ read $\sigma(t)=(t,q(t))$ and $\rho(t)=(t,m(t))$, where $m(t)=\pi_{M,Q}(q(t))\in M$. The first jet extensions $j^1\sigma$ and $j^1\rho$ are identified with $(t,\dot q(t))$ and $(t,\dot m(t))$.

In this particular situation, the connection $\mathcal{A}$ on $P$ is always chosen to be induced by a connection $\gamma$ on $Q$. In this case, the reduced section $\bar\sigma$ is identified with $(t,\bar v(t))$, where $\bar v(t)=[q(t),\gamma(\dot q(t))]_G$. Similarly, a section $\xi\in\Gamma(\pi_{\mathbb{R},\operatorname{Ad}P})$ covering $\rho(t)=(t,m(t))$ reads $\xi(t)=(t,\bar\xi(t))$, where $\bar\xi(t)\in \operatorname{Ad}_{m(t)}Q$. The $\nabla^\mathcal{A}$-derivative $\tilde\nabla^\mathcal{A}$ of $\xi$ can thus be identified with the covariant time derivative $\lp D^\gamma/Dt\rp\bar\xi(t)$. Using all these observations, the second equation of \eqref{LP_field_eqn} reads
\[
\frac{D^\gamma}{Dt}\frac{\delta l}{\delta\bar v}-\operatorname{ad}^*_{\bar v}\frac{\delta l}{\delta\bar v}=0
\]
and the variation of $\bar v$ is $\delta^\gamma\bar v=\frac{D^\gamma}{dt}\bar\eta-[\bar\eta,\bar v]+\tilde{B}(\delta m,\dot m)$, where $\tilde{B}$ is the reduced curvature of $\gamma$.
Recall that writing the horizontal equation requires a projectable covariant derivative $\nabla^\Sigma$ on $T\Sigma$, which is also assumed to be torsion free for simplicity. In this classical case, the covariant derivative is constructed from a torsion free covariant derivative $\nabla$ on $TM$ and the natural covariant derivative on $T\mathbb{R}$. In this case, $\nabla^\Sigma$ is obviously projectable and torsion free. The first equation of \eqref{LP_field_eqn} reads
\[
\frac{\delta l}{\delta m}-\frac{D^\nabla}{Dt}\frac{\delta l}{\delta \dot m}=\left\langle\frac{\delta l}{\delta\bar v},\mathbf{i}_{\dot m}\tilde{\mathcal{B}}\right\rangle.
\]
Thus the classical Lagrange-Poincar\'e equations obtained by standard Lagrangian reduction are recovered; see Theorem 3.4.1 in \cite{CeMaRa2001}. Note that here the Lagrangian is allowed to be time-dependent.

If $\nabla$ has a torsion $T^\nabla$, the horizontal equation reads 
\[
\frac{\delta l}{\delta m}-\frac{D^\nabla}{Dt}\frac{\delta l}{\delta \dot m}+\left\langle\frac{\delta l}{\delta \dot m}, \mathbf{i}_{\dot m}T^\nabla\right\rangle=\left\langle\frac{\delta l}{\delta\bar v},\mathbf{i}_{\dot m}\tilde{\mathcal{B}}\right\rangle,
\]
see \eqref{LP_eqn_torsion}.

In the particular case, $G=\{1\}$, there is no reduction and the vertical equation is absent. In this case the horizontal equation reads
\[
\frac{D^\nabla}{Dt}\frac{\delta L}{\delta \dot q}-\frac{\delta L}{\delta q}=0.
\]
Of course, this recovers the standard Euler-Lagrange equation written with the help of a connection. In the case when the connection has torsion, this reads
\[
\frac{D^\nabla}{Dt}\frac{\delta L}{\delta \dot q}-\frac{\delta L}{\delta q}=\left\langle\frac{\delta l}{\delta \dot q}, \mathbf{i}_{\dot q}T^\nabla\right\rangle;
\]
see (3) in \cite{GaSo2003}. Recall that the usual way to write the Euler-Lagrange equations
\[
\frac{d}{dt}\frac{\partial L}{\partial \dot q}-\frac{\partial L}{\partial q}=0
\]
makes sense only locally; see \eqref{EL_local}.

Another particular case arises when $Q=G$. In this case, there is no horizontal equation and the vertical equation gives the Euler-Poincar\'e equation. Indeed, in this case, all the connections are equivalent (the bundle $Q \rightarrow Q/G $ is over a point) and the covariant time derivative on the adjoint bundle becomes the ordinary time derivative on the Lie algebra $\mathfrak{g}$. These observations and \eqref{LP_field_eqn} recover the Euler-Poincar\'e equation together with the associated constrained variations
\[
\frac{d}{dt}\frac{\delta l}{\delta v}-\operatorname{ad}^*_v\frac{\delta l}{\delta v}=0\quad\text{and}\quad \delta v=\frac{d}{dt}\eta-[\eta,v].
\]
\end{itemize}

\subsection{Reconstruction}\label{sec:Reconstruction}

Having derived the Lagrange-Poincar\'e field equations it is natural to turn to the problem of reconstruction of solutions to the original Euler-Lagrange equation from solutions to the reduced equations. More precisely, given a solution section $\bar\sigma:\bar{U}\rightarrow L(\pi_{X,\Sigma}^*TX,\operatorname{Ad}P)$ of the Lagrange-Poincar\'e equations, how can one construct a solution section $\sigma:\bar{U}\rightarrow P$ of the Euler-Lagrange equations? Note that Theorem \ref{EP_field_reduction} does not consider this problem, since the section $\sigma$ is given a priori.  This section deals with the reconstruction problem and demonstrates that reconstruction of field theories requires an extra integrability condition.

\medskip

\paragraph{Induced connection.} A section $\bar\sigma\in\Gamma(\pi_{X,L(TX,\operatorname{Ad}P)})$ induces a $G$-principal bundle $P^\rho\rightarrow X$ and a connection $\mathcal{A}^{\bar\sigma}$ on it as follows: The subset $P^\rho\subset P$ is defined by
\[
P^\rho:=\pi_{\Sigma,P}^{-1}(\rho(X))=\left\{p\in P\mid \pi_{\Sigma,P}(p)\in\rho(X)\right\},
\]
where $\rho:=\pi_{\Sigma,L\left(\pi_{X,\Sigma}^*TX,\operatorname{Ad}P\right)}\circ \bar\sigma\in \Gamma(\pi_{X,\Sigma})$. Since $\rho$ is a section, it is an injective immersion and a homeomorphism onto its image. Thus, the image $\rho(X)$ is a submanifold of $\Sigma$. Now, since $\pi_{\Sigma,P}$ is a submersion, it is transversal to the submanifold $\rho(X)$. This proves that $P^\rho$ is a submanifold of $P$, whose tangent space at $p$ is
\[
T_pP^\rho=\left(T_p\pi_{\Sigma,P}\right)^{-1}(T_s\rho(T_s\Sigma)),\quad s=\pi_{\Sigma,P}(p).
\]
The manifold $P^\rho$ may be endowed with the structure of a $G$-principal bundle over $X$ by restriction of the $G$-action on $P$. Note that $P^\rho$ can be identified with the pull-back bundle $\rho^*P=\{(x,p)\mid \pi_{\Sigma,P}(p)=\rho(x)\}$, the identification being given by
\[
p\in P^\rho \mapsto (\pi_{X,P}(p),p)\in \rho^*P.
\]
The section $\bar\sigma$ may be regarded as a section of the vector bundle $L(TX,\rho^*\operatorname{Ad}P)\simeq L(TX,\operatorname{Ad}P^\rho)$, and thus induces an equivariant and vertical one-form $\omega^{\bar\sigma}\in\overline{\Omega^1}(P^\rho,\mathfrak{g})$.  The isomorphism $\overline{\Omega^1}(P^\rho, \mathfrak{g}) \leftrightarrow L(TX, \operatorname{Ad} P^ \rho) $ is written explicitly as follows:
\begin{equation}\label{def_iso_bar}
\omega^{\bar{ \sigma}} \in \overline{\Omega^1}(P^\rho, \mathfrak{g}) \mapsto \bar{\sigma} \in L(TX, \operatorname{Ad} P^ \rho), \quad 
\bar{\sigma}(u_x) : = \llsb p, \omega^{\bar{ \sigma}}(u_p) \rrsb_{\mg},
\end{equation}
where $u_x = T\pi_{X,P}(u_p) \in T_xX$, $u_p \in T_pP$. 
The connection $\mathcal{A}$ on $\pi_{\Sigma,P}:P\rightarrow\Sigma$ naturally induces a connection $\mathcal{A}^\rho$ on $P^\rho\rightarrow X$.  A new connection $\mathcal{A}^{\bar\sigma}$ is thereby obtained on $P^\rho\rightarrow X$.  Concretely,
\[
\mathcal{A}^{\bar\sigma}:=\mathcal{A}^\rho-\omega^{\bar\sigma}.
\]
Thus one may interpret the vertical solution of the Lagrange-Poincar\'e field equations as describing an affine modification to the a priori connection $\cA^\rho$.  The modified connection $\cA^{\bar\sig}$ is the correct choice of connection for reconstruction, as is explained below.

\paragraph{Reconstruction condition.} We now prove that if $\bar\sigma$ is the reduced section associated to a section $\sigma\in\Gamma(\pi_{X,P})$ then $\mathcal{A}^{\bar\sigma}$ is flat. Indeed, in this case $P^\rho=\{\Phi_g(\sigma(x))\mid g\in G, x\in X\}$ and for $p=\sigma(x)\in P^\rho$ and $v_p\in T_pP^\rho$ formula \eqref{def_iso_bar} gives 
\begin{equation}\label{horizontal_condition}
\mathcal{A}^{\bar\sigma}(v_p)=\mathcal{A}^\rho(v_p)-\omega^{\bar\sigma}(v_p)=\mathcal{A}(v_p)-\mathcal{A}(T_x\sigma(T_p\pi_{X,P}(v_p)))
\end{equation}
since $\bar {\sigma}(x) = \llsb \sigma(x), \sig^*\mathcal{A}(x)\rrsb_{\mg}$ for all $x \in X $. Recall that $u_p\in T_pP^\rho$ if and only if $T_p\pi_{\Sigma,P}(u_p)\in T_x\rho(T_xX)$. That is, in terms of $\sigma$,
\[
T_p\pi_{\Sigma,P}(u_p)\in T_{\sigma(x)}\pi_{\Sigma,P}\left(T_x\sigma(T_xX)\right).
\]
This proves that $T_{\sigma(x)}P^\rho=T_x\sigma(T_xX)+V_{\sigma(x)}P$ at $p=\sigma(x)$,
where $V_{\sigma(x)}P=\operatorname{ker}(T_{\sigma(x)}\pi_{\Sigma,P})$ is the vertical space relative to $\pi_{\Sigma,P}$. Thus, for $p=\sigma(x)$, any $v_p\in T_pP^\rho$ reads $v_p=T_x\sigma(v_x)+\xi_P(p)$. Inserting this expression for $v_p$ into \eqref{horizontal_condition}, reveals the condition $\xi=0$. This proves that the $\mathcal{A}^{\bar{ \sigma}}$-horizontal subspace at $\sigma(x)$ is given by
\[
H^{\mathcal{A}^{\bar{ \sigma}}}_{\sigma(x)}P=T_x\sigma(T_xX).
\]
This horizontal distribution is integrable, the integral leaves being given by $\{\Phi_g(\sigma(x))\mid x\in X\}=\Phi_g(\operatorname{Im}(\sigma))$, for each $g \in G $. Thus, the connection $\mathcal{A}^{\bar\sigma}$ on $P^\rho$ is flat and the \textit{horizontality condition}
\begin{equation}\label{horizontality_condition}\sigma ^\ast\mathcal{A}^{\bar{ \sigma}} = 0.\end{equation} 
is a necessary condition for reconstruction.

Conversely, consider a section $\bar\sigma$ of $\pi_{X,L(TX,\operatorname{Ad}P)}$ such that the connection $\mathcal{A}^{\bar\sigma}$ on $P^\rho$ is flat and has trivial holonomy. Since the connection $\mathcal{A}^{\bar\sigma}$ is flat, the horizontal distribution is integrable and the leaves cover the base, that is, given a leaf $L$, each fiber intersects the leaf $L$ at least once. Since the holonomy is trivial, each fiber intersects the leaf exactly once. This construction shows that each integral leaf of the horizontal distribution defines a section of the bundle $P^\rho\rightarrow X$.  Thus a family of sections $\Phi_g\circ\sigma$ of $\pi_{X,P}$ that project via $\pi_{\Sigma,P}$ to $\rho$ is attained.
Since
\[
\llsb \sigma,\sig^*\mathcal{A}\rrsb_{\mg}=\llsb\sigma,\sig^*\mathcal{A}^{\bar\sigma}+\omega^{\bar\sigma}\rrsb_{\mg}=\llsb\sigma,\omega^{\bar\sigma}\rrsb_{\mg}=\bar\sigma,
\]
the section $\bar\sigma$ is the reduced section associated to the family of sections $\Phi_g\circ\sigma$ for each $g  \in G $. The horizontality condition \eqref{horizontality_condition} is, of course, satisfied.

Recall that the flatness of the connection does not imply that the holonomy is trivial unless the base is simply connected or the holonomy group is connected. Note that this fact implies that the holonomy of a flat connection is locally trivial, that is, for every $x\in X$, there exists an open neighborhood $U$ such that the holonomy of $P|_U$ is trivial.

The situation is summarized in the following reconstruction theorem.

\begin{theorem}\label{ReconstructionTheorem} Fix a connection $\mathcal{A}$ on the principal bundle $\pi_{\Sigma,P}:P\rightarrow \Sigma$, consider a $G$-invariant Lagrangian $L$ and the reduced Lagrangian $l$.

If $\sigma:\bar{U}\rightarrow P$ is a solution of the Euler-Lagrange field equations, then the reduced section $\bar\sigma$ is a solution of the Lagrange-Poincar\'e field equations. Moreover the connection $\mathcal{A}^{\bar\sigma}$ on $P^\rho$ is flat and the horizontality condition \eqref{horizontality_condition} holds.

Conversely, given a solution $\bar\sigma$ of the Lagrange-Poincar\'e equations on $\bar{U}$ such that $\mathcal{A}^{\bar\sigma}$ is flat and has trivial holonomy over an open set containing $\bar{U}$, the family $\Phi_g\circ \sigma$, $g\in G$, of solutions of the Euler-Lagrange field equations are given by the integral leaves of the horizontal distribution associated to $\mathcal{A}^{\bar\sigma}$. In addition, the horizontality condition \eqref{horizontality_condition} holds. If the connection $\mathcal{A}^{\bar\sigma}$ is flat one can always restrict it to an open simply connected set contained in $U$ so that its holonomy on $U $ is automatically zero.
\end{theorem}

Note that the curvature of $\mathcal{A}^{\bar\sigma}$ is $\mathcal{B}-\mathbf{d}^\mathcal{A}\omega^{\bar\sigma}-\lp 1/2\rp\omega^{\bar\sigma}\wedge\omega^{\bar\sigma}$. Therefore, the \textit{reconstruction condition} is
\begin{equation}\label{reconstruction} \mathcal{B}-\mathbf{d}^\mathcal{A}\omega^{\bar\sigma}-\half\omega^{\bar\sigma}\wedge\omega^{\bar\sigma}=0\quad\text{on}\quad P^\rho.
\end{equation}
This condition has to be seen as an equality in the space $\overline{\Omega^2}(P^\rho,\mathfrak{g})$ of equivariant vertical two-forms. The isomorphism \eqref{def_iso_bar} shows it is equivalent to assume that the corresponding two-form in $\Omega^2(X,\operatorname{Ad}P^\rho)=\Omega^2(X,\rho^*\operatorname{Ad}P)$ vanishes. Applying \eqref{def_iso_bar} to equation \eqref{reconstruction} recovers the formula
\begin{equation}\label{reconstruction1} 
d^\cA\bar\sig+\half\bar\sig\wedge\bar\sig=\rho^*\tilde{\mathcal{B}}.
\end{equation}

\paragraph{Reconstruction equation.}
When reconstructing solutions of the Euler-Lagrange field equations one needs 
to add \eqref{reconstruction1} to the reduced field equations \eqref{LP_field_eqn} since there could be solutions to the Lagrange-Poincar\'e field equations \eqref{LP_field_eqn} that do not correspond to the original Euler-Lagrange system.  Given a solution $\lp \rho, \bar\sig\rp$ as specified above, \eqref{def_iso_bar} uniquely determines $\om^{\bar\sig}$ by the formula
\begin{equation}\label{VerticalReconstructionForm}
\bar\sig(u_x) = \left [ \!\! \left [ p,\om^{\bar\sig}\left(\hor_p^{\cA}(T_x\rho(u_x)) \right)\right ] \!\! \right ]_{\mg}, \qquad p \in \pi_{\Sig,P}^{-1}\lp \rho(x)\rp, \qquad u _x\in T _xX,
\end{equation}
since $T_p \pi_{X, P} \left(\operatorname{Hor}_p ^ \mathcal{A}(T_x \rho(u_x)) \right) = u_x$. Thus, $\omega^{\bar{ \sigma}}$ is completely determined in terms of $( \rho, \bar{ \sigma})$.

For a section $\sig \in \Gamma\lp \pi_{X, P^\rho}\rp$, the horizontality condition \eqref{horizontality_condition} for $\cA^{\bar\sig}$ is
\begin{equation}\label{HorizontalityCondition}
0 = \sig^*\cA^{\bar\sig} = \sigma^\ast \mathcal{A}^ \rho - \sigma ^\ast \omega^{ \bar{ \sigma}} =  \sig^*\cA^ \rho - \rho^*\lp \hor^{\cA}_ \sigma\rp^*\om^{\bar\sig} 
\end{equation}
because $\omega^{ \bar{ \sigma}}\left(T_x \sigma( u _x) \right) = \omega^{ \bar{ \sigma}}\left(\operatorname{Hor}^ \mathcal{A}_ { \sigma(x)}\left(T_x \rho( u _x) \right) \right)$ since $\omega^{ \bar{ \sigma}} \in 
\overline { \Omega^1} (P^ \rho, \mathfrak{g})$.
Note that following the determination of $\om^{\bar\sig}$ by \eqref{VerticalReconstructionForm} the only unknown quantity in \eqref{HorizontalityCondition} is $\sig$.  We now show that  \eqref{HorizontalityCondition} gives a first order PDE that determines $\sigma$.

If $u _x\in T _xX$, then by \eqref{HorizontalityCondition} and the horizontal-vertical decomposition relative to the connection $\mathcal{A}^ \rho$,
\begin{align*}
T _x\sigma ( u _x) & = \operatorname{Hor}^{\mathcal{A}^ \rho}_{ \sigma (x)} \left(T_{ \sigma(x)} \pi_{ \Sigma, P} \left(T _x \sigma ( u _x) \right) \right) + 
\left(\mathcal{A}^ \rho \left(T _x\sigma( u _x) \right) \right)_P (\sigma (x))\\ 
& = \operatorname{Hor}^{\mathcal{A}^ \rho}_{ \sigma (x)} \left(T_x \rho (u _x) \right) + \left(\sigma^\ast \mathcal{A}^ \rho ( u _x) \right)_P ( \sigma(x)) \\
& = \operatorname{Hor}^{\mathcal{A}^ \rho}_{ \sigma (x)} \left(T_x \rho (u _x) \right) + \left(\rho ^\ast \left(\operatorname{Hor}_ \sigma^{\mathcal{A}^ \rho} \right) ^\ast \omega^{\bar{\sigma}}\right)_P ( \sigma(x)).
\end{align*}
This gives the following first order reconstruction PDE for $\sigma$:
\begin{equation}
T_x\sig = \hor^{\cA}_{\sig(x)}\circ T_x\rho + \lp \om^{\bar\sig}\circ \hor^{{\cA}^ \rho}_{\sig(x)}\circ T_x\rho\rp_P(\sig(x)).
\label{eq:ExplicitReconstruction}
\end{equation}

Theorem \ref{ReconstructionTheorem} can now be interpreted as asserting that given a solution $(\rho,\bar\sig)$ of equations \eqref{LP_field_eqn} and \eqref{reconstruction1}, \textit{there exists a unique solution} $\sig$ to the reconstruction equation \eqref{eq:ExplicitReconstruction} in a neighborhood where $\cA^{\bar\sig}$ has trivial holonomy. This section $\sigma$ solves the corresponding Euler-Lagrange equations for the unreduced problem.

As a final comment, note that \eqref{eq:ExplicitReconstruction} is the field theoretic analogue of the classical reconstruction equation $\xi = g ^{-1} \dot{g} $ associated to the Euler-Poincar\'e equations.

\paragraph{Particular cases.}
The reconstruction condition specializes to the particular cases as follows:
\begin{itemize}
\item[{\bf i}] It $G=\{1\}$ there is no reduction and, therefore, no reconstruction condition.
\item[{\bf ii}] In this case the variable $\rho$ is absent, so $P^\rho=P$. Moreover, the reduced section $\bar\sigma$ turns out to be associated, via the map $\beta_\mathcal{A}$, to a section $\zeta$ of $J^1P/G\rightarrow X$, that can be interpreted as a connection on $P$. This connection $\zeta$ does not depend on the chosen $\mathcal{A}$ and $\mathcal{A}^{\bar\sigma}$ turns out to be the connection one-form associated to $\zeta$. The reconstruction condition is simply that the curvature of this connection (or of $\zeta$) is zero.  This recovers the reconstruction condition that in the case of covariant Euler-Poincar\'e reduction; see \S 3.2 of \cite{CaRaSh2000}.
\item[{\bf iii}] The reconstruction condition is the same as in \cite{CaRa2003}.
\item[{\bf iv}] Since $X=\mathbb{R}$, the base is one-dimensional and every connection is flat. Since $\mathbb{R}$ is simply connected the holonomy is trivial. The reconstruction condition is always satisfied. This agree with the fact that in classical Lagrangian reduction, the solution of the Euler-Lagrange equations can always be constructed from that of the reduced equations.
\end{itemize}

\section{Conservation laws and representations}\label{cons_law}

In applications there is often a natural choice of gauge that is used to formulate the Lagrange-Poincar\'e field equations in a convenient local form.

This section describes the two predominant choices of representations that occur, the {\it spatial representation} and the {\it convective representation}.  The Lagrange-Poincar\'e equations \eqref{LP_field_eqn} are given locally using these choices of gauge.  The spatial representation yields Noether's Theorem as the vertical equation whilst the convective representation has the Euler-Poincar\'e equation as its vertical equation.  This observation shows that  the Lagrange-Poincar\'e equations are equivalent to Noether's Theorem, a statement often found in the literature when dealing with concrete applications.

This section also formulates a global version of the Kelvin-Noether Theorem that generalizes the result for classical systems given, for example, in \cite{CeHoMaRa1998,HoMaRa1998}.

\subsection{Representations and Noether's Theorem}\label{sec:Representations}

A section $\sigma \in \Gamma\lp \pi_{X,P}\rp$ introduces a representation of $\Ad^*\,P$ which, in turn, yields local equations for the vertical part of \eqref{LP_field_eqn}.  The two natural choices of section and their associated representations are described below.

\paragraph{Convective representation.} Suppose one seeks a local solution $\sig:U\subset X\rightarrow P$ of \eqref{LP_field_eqn} and \eqref{eq:ExplicitReconstruction} in a trivialization of $P$ over $U\subset X$.  Let $\rho := \pi_{\Sig, P}\circ \sig$ and $V:=\rho\lp U\rp\subset \Sig$.  Suppose further that a flat connection $\mathcal{A}$ exists on $P\rightarrow V$. Then, there exists a unique section $\gamma:V\rightarrow P$ such that $T_s\gamma(v_s)\in H_sP$, for all $s\in \Sigma$ and $v_s \in T_s\Sigma$.  Therefore, the section $\sigma^h:=\gamma\circ\rho\in \Gamma(\pi_{U,P})$ has the property that $T_x\sig^h\lp v_x\rp \in H_{\sig^h(x)}P$ for all $v_x \in TX|_U$. Such a section is called a {\em horizontal section}.

\begin{remark}\normalfont It may not be possible to find a flat connection $\cA$ on an arbitrary open set $V\subset \Sig$.  The convective representation is not defined in such cases.  In applications one may find that shrinking the set $U\subset X$ yields a suitable $V\subset \Sig$ such that the convective representation makes sense.
\end{remark}

The map $g:U\subset X\to G$ such that $\sig(x) = \Phi_{g(x)}\sig^h(x)$ for all $x\in U$ together with \eqref{eq:Rep} produce
\[
\chi_{\sig^h}\lp \bar\sig\rp = \chi_{\sig^h}\lp\llsb \sig, \sig^*\cA\rrsb_{\mg}\rp = g^{-1}dg =:\xi \in \Om^1\lp X,\mg\rp.
\]
Consequently, the vertical part of equations \eqref{LP_field_eqn} composed with $\chi_{\sigma^h}$ yields
\[
\chi_{\sig^h}\lp\dive^{\cA}\dede{l}{\bar \sig} - \ad^*_{\bar\sig}\dede{l}{\bar\sig}\rp=\chi_{\sig^h} \lp\left [ \!\!\left[ \sig^h, \operatorname{div}\dede{l}{\xi} - \ad^*_{\xi}\dede{l}{\xi}\right ] \!\!\right]_{\mg^{*}}\rp= \dive\dede{l}{\xi} - \ad^*_{\xi}\dede{l}{\xi}.
\]
Thus, the local representation of the vertical Lagrange-Poincar\'e equation in this gauge is
\begin{equation}
\dive\dede{l}{\xi} - \ad^*_{\xi}\dede{l}{\xi} = 0
\label{eq:EP}
\end{equation}
which recovers the Euler-Poincar\'e equation.  This choice of gauge is called the \textit{convective representation}, see \cite{CeHoMaRa1998}.

\paragraph{Spatial representation.}
With the same notation as for the convective representation, $\sig = g\sig^h$, the map $\chi_{\sig}$ applied to $\bar \sig$ yields
\[
\chi_{\sig}\lp \bar\sig\rp = \chi_{\sig}\lp\llsb \sig, \sig^*\cA\rrsb_{\mg}\rp = dgg^{-1} =:\Xi \in \Om^1\lp X,\mg\rp.
\]
Accordingly,
\[
\chi_{\sigma}\left(\dive^{\cA}\dede{l}{\bar \sig} - \ad^*_{\bar\sig}\dede{l}{\bar\sig}\right)=\chi_{\sigma}\left(\left [ \!\!\left[ \sig, \operatorname{div}\dede{l}{\Xi}\right ] \!\!\right]_{\mg^{*}}\right)=\operatorname{div}\dede{l}{\Xi}.
\]
Thus, the local representation of the vertical Lagrange-Poincar\'e equation \eqref{LP_field_eqn} in this gauge reads
\begin{equation}
\dive\dede{l}{\Xi} = 0
\label{eq:Noether}
\end{equation}
which is Noether's Theorem.  This choice of gauge is called the \textit{spatial representation}.  Note that $\Ad_g\xi = \Xi$.  Therefore both the Euler-Poincar\'e equations and Noether's Theorem are local representations of equations \eqref{LP_field_eqn} corresponding to a particular choice of gauge.  In particular, the Euler-Poincar\'e equation is equivalent to Noether's Theorem.

\begin{remark}\normalfont When the convective representation can not be defined it is still possible to fix $\Xi = \sig^*\cA$ and proceed with the construction of the spatial representation without the use of $\sig^h$.  Thus the spatial representation is always well-defined, while the convective representation is not.
\end{remark}

\begin{remark}\normalfont  In classical Lagrangian reduction when $X=\mR$ it is always possible to construct a local horizontal section.  Therefore the convective representation is always well-defined for classical systems. 
\end{remark}

\subsection{The Kelvin-Noether theorem}

Given any manifold $\cC$ on which $G$ acts, the \textit{associated bundle} is a fiber bundle over $\Sig$ defined by
\[
\cH_{\cC} := P\times_G \cC = \lp P \times \cC\rp/G,
\]
where the action of $G$ on $P\times \cC$ is the diagonal action.  The adjoint and coadjoint bundles, $\Ad\,P$ and $\Ad^*\,P$ are associated bundles with $\cC = \mg$ and $\cC = \mg^*$ respectively.  The action of $G$ on $\mg$ for $\Ad\,P$ is the adjoint action whilst the action on $G$ on $\mg^*$ is the coadjoint action.  The equivalence class of $(p,c) \in P\times \cC$ is denoted
\[
\llsb p, c\rrsb_{\cC} \in P\times_G\cC.
\]
The lifted action of $G$on $T\cC$ enables the definition of $\cH_{T\cC} = P\times_G T\cC$.  The infinitesimal action $\Ad\,P\times \cH_{\cC} \to \cH_{T\cC}$ on $\cH_{\cC}$ as follows:
\begin{equation}
\llsb p, \xi\rrsb_{\mg} \cdot \llsb p, c\rrsb_{\cC} = \llsb p, \xi_{\cC}\lp c\rp \rrsb_{T\cC},
\label{eq:InfAc}
\end{equation}
where the vector field $\xi_\mathcal{C}\in\mathfrak{X}(\mathcal{C})$ denotes the infinitesimal generator of $\xi\in\mathfrak{g}$ on $\mathcal{C}$.

A connection form $\cA$ on $\pi_{\Sigma,P}$ yields a {\em covariant tangent functor} $T^{\cA}$ defined on sections of $\pi_{X, \cH_{\cC}}$ by
\begin{equation}
T^\cA: \Gamma\lp \pi_{X, \cH_{\cC}}\rp \to \Gamma\lp \pi_{X,L(\pi_{X,\Sigma}^*TX, \cH_{T\cC})}\rp, \qquad  T^\cA\llsb \sig, c\rrsb_{\cC} = \llsb \sig , Tc - \lp \sig^*\cA\rp_{\cC}\circ c \rrsb_{T\cC}.
\label{d_A_associated}
\end{equation}
Therefore, if $\bar c$ is a section of $\pi_{X,\mathcal{H}_\mathcal{C}}$ covering the section $\rho$ of $\pi_{X,\Sigma}$, then $T^\mathcal{A}\bar c(v_x)\in (\mathcal{H}_{T\mathcal{C}})_{\rho(x)}$.

The map $K$ can be described by a $G$-equivariant map $\cK:\cC \to \mg^{**}$ defined by the relation
\[
K\lp \llsb p, c\rrsb_{\cC}\rp = \llsb p, \cK\lp c\rp\rrsb_{\mg^{**}}.
\]
Here $\mg^{**}$ denotes the double dual of the Lie algebra.  For an example where the distinction between $\mg$ and $\mg^{**}$ arises, see \cite{HoMaRa1998}.  The derivative of $K:\cH_{\cC} \to \Ad^{**}\,Q$ may be defined as follows:
\[
dK\lp \llsb p, v_c\rrsb_{\cC}\rp=\lp \llsb p, \mathbf{d}\mathcal{K}(v_c)\rrsb_{\cC}\rp.
\]
Note that $dK:(\mathcal{H}_{T\mathcal{C}})_s\rightarrow (\operatorname{Ad}^{**}P)_s$ and $d^{\cA}\lp K\circ \bar c\rp = dK\circ T^{\cA}\bar c$, where $\bar c=\llsb\sigma,c\rrsb_\mathcal{C}$ denotes a section of $\pi_{X,\mathcal{H}_\mathcal{C}}$. Indeed,
\begin{align*}
d^{\cA}\lp K\circ \bar c\rp &= \llsb \sig, d\lp \cK\circ c\rp - \lp \sig^*\cA\rp \cdot \lp \cK\circ c\rp\rrsb_{\mg^{**}}\\
&= \llsb \sig, d\cK\circ Tc - d\cK\circ \lp\sig^*\cA\rp_{\cC}\circ c  \rrsb_{\mg^{**}}\\
&= \llsb \sig, d\cK\circ\lp Tc - \lp\sig^*\cA\rp_{\cC}\circ c\rp\rrsb_{\mg^{**}}\\
&= dK\circ T^{\cA}\bar c.
\end{align*}
This relation is described by the following commutative diagram:
\begin{diagram}
\Gamma\lp \pi_{X,L(\pi_{X,\Sigma}^*TX, \cH_{T\cC})}\rp & \rTo^{dK} & \Gamma\lp \pi_{X,L(\pi_{X,\Sigma}^*TX, \Ad^{**}\,P)}\rp\\
\uTo^{T^\cA} & & \uTo_{d^{\cA}}\\
\Gamma(\pi_{X,\cH_{\cC}}) & \rTo_{K} & \Gamma(\pi_{X,\Ad^{**}\,P})\\
\end{diagram}
The infinitesimal actions described in \eqref{eq:InfAc} on $\Ad^{**}\,P$ and $\cH_\cC$ are readily observed to be related via
\begin{equation}
\llsb p, \xi\rrsb_{\mg} \cdot K\lp \bar c\rp = dK \lp \llsb p, \xi\rrsb_{\mg}\cdot \bar c\rp.
\label{eq:InfAcRel}
\end{equation}
Additionally, observe the following relationship:
\begin{equation}\label{eq:cov2}
\operatorname{div}\scp{\bar\nu}{\bar \mu} = \scp{d^{\cA}\bar\nu}{\bar \mu} + \scp{\bar\nu}{\dive^{\cA}\bar \mu},
\end{equation}
where $\bar \nu$ and $\bar\mu$ are sections of $\pi_{X,\operatorname{Ad}^{**}P}$ and $\pi_{X,L(\pi^*_{X,\Sigma}T^*X,\operatorname{Ad}P^*)}$ respectively, and both cover the same section $\rho$ of $\pi_{X,\Sigma}$.

The Kelvin-Noether Theorem may be stated as follows:

\begin{theorem}\label{thm:Kelvin-Noether}  Let $\lp \rho, \bar\sig\rp \in \Gamma\lp \pi_{X,\Sig}\rp \times \Om^1\lp X,\Ad\,P\rp$ be a solution to the Lagrange-Poincar\'e equations \eqref{LP_field_eqn}, and $\bar c \in \Gamma\lp \pi_{X, \cH_{\cC}}\rp$ cover $\rho$ while satisfying
\begin{equation}
T^\cA\bar c + \bar\sig_{\cC}\circ \bar c = 0.
\label{eq:Kel-NoCondition}
\end{equation}
If $K: \cH_{\cC} \to \Ad^{**}P$ fiber-preserving map that covers the identity on $\Sigma$ then the associated circulation
\[
I:= \scp{K\circ \bar c}{\dede{l}{\bar \sig}} \in \mathfrak{X}(X)
\]
satisfies
\begin{equation}
\operatorname{div}I = 0.
\label{eq:Kel-No}
\end{equation}
\end{theorem}
\begin{proof}  The result is obtained via a direct calculation that uses \eqref{LP_field_eqn}, \eqref{eq:cov2}, \eqref{eq:InfAcRel}, and \eqref{eq:Kel-NoCondition} as follows:
\begin{align*}
\operatorname{div}I &= \operatorname{div}\scp{K\circ \bar c}{\dede{l}{\bar \sig}}\\
&= \scp{d^\cA\lp K\circ \bar c\rp}{\dede{l}{\bar \sig}} + \scp{K\circ \bar c}{\dive^{\cA}\dede{l}{\bar \sig}}\\
&= \scp{\lp d^\cA + \ad^{**}_{\bar\sig}\rp\lp K\circ \bar c\rp}{\dede{l}{\bar \sig}} + \scp{K\circ \bar c}{\lp \dive^{\cA} - \ad^*_{\bar\sig}\rp\dede{l}{\bar \sig}}\\
&= \scp{\lp d^\cA + \ad^{**}_{\bar\sig}\rp\lp K\circ \bar c\rp}{\dede{l}{\bar \sig}}\\
&= \scp{ dK\circ \lp T^{\cA}\bar c + \bar\sig_{\cC}\circ \bar c\rp}{\dede{l}{\bar \sig}} = 0,
\end{align*}
as required.
\end{proof}

\medskip

Recall that classical Lagrangian reduction (particular case {\bf iv}), used the formulation $\sig(t) = (t, q(t))$ and $\bar \sig(t) = \llsb q(t), \cA\lp \dot q(t)\rp\rrsb_{\mg}$.  In this case \eqref{eq:Kel-NoCondition} becomes
\[
T^{\cA} \bar c + \bar \sig_{\cC}\circ \bar c = \llsb q, \dot c - \lp\cA\lp \dot q\rp\rp_{\cC}\circ c\rrsb_{T\cC} + \llsb q, \lp\cA\lp \dot q\rp\rp_{\cC}\circ c \rrsb_{T\cC}=  \llsb q, \dot c\rrsb_{T\cC}= 0.
\]
Therefore \eqref{eq:Kel-NoCondition} diminishes to the assumptions for the classical Kelvin-Noether theorem; see \cite{HoMaRa1998}.  Furthermore the conclusion to Theorem \ref{thm:Kelvin-Noether} in this context becomes
\[
\dd{}{t}\scp{\cK\lp c\rp}{\dede{l}{\xi}} = 0.
\]
These results extend those of \cite{CeHoMaRa1998} to the Lagrange-Poincar\'e context.

\section{Applications}\label{sec:Applications}

This Section presents a brief outline of some applications of the Lagrange-Poincar\'e field equations.

The first application is the minimal immersion problem, which is treated explicitly in local coordinate form. 

The second and third applications are classical and covariant metamorphosis image dynamics treated in coordinate-free form.  The classical formulation may be understood as a change of variables from the treatment given in \cite{HoTrYo2008}.  This coordinate transformation formally decouples the equations. The covariant formulation provides a new insight into the problem.  This covariant formulation aims to provide a basis for the future use of multisymplectic integrators in image dynamics.

\subsection{Minimal immersions}\label{sec:MinImm}

An interesting classical problem that has applications ranging from the shape of soap bubbles to string theory is that of minimal embeddings or, more generally, minimal immersions.  In string theory the Nambu action describes a world sheet in spacetime; see for example \cite{Nak2003}.  The soap bubble problem is a generalization of the isoperimetric problem studied by Newton amongst others.  

Given a manifold $X$ and a pseudo-Riemannian manifold $(Q,g)$, the problem is to find an immersion $\eta \in \operatorname{Imm}(X,Q)$ such that the surface area of $\eta\lp X\rp$ is minimized.  For the soap bubble problem there is an additional constraint on the volume enclosed by $\eta\lp X\rp$ which we shall not treat here.

The minimal immersion problem may be cast into the bundle picture described in \S\ref{sec:geometric_constructions} as follows:  Let $P= X \times Q$ and $\pi_{X,P}=p_1$ be projection on the first factor.  Since $\pi_{X,P}$ is a trivial fiber bundle, sections $\sig \in \Gamma\lp \pi_{X,P}\rp$ of $\pi_{X,P}$ can be represented by smooth maps $\eta:X \to Q$, namely $\sigma( x) = ( x , \eta (x)) $.  Further, consider sections $\sig \in \Gamma\lp \pi_{X,P}\rp$ such that $\eta \in Imm\lp X,Q\rp$ is an immersion, that is, $\sig \in \Gamma^{Imm}\lp \pi_{X,P}\rp$ where
\[
\Gamma^{\operatorname{Imm}}\lp \pi_{X,P}\rp = \left\{ \sigma \in \Gamma\lp \pi_{X,P}\rp \mid \eta \in \operatorname{Imm}\lp X,Q\rp \right\}.
\]
Since $\eta$ is an immersion, $h=\eta^*g$ is a metric on $X$. Locally $h$ reads
\[
h_{ij}=(g_{\alpha\beta}\circ\eta)\partial_i\eta^\alpha\partial_j\eta^\beta,
\]
where the indices $i,j,...$ denotes coordinates on $X$ and $\alpha,\beta,...$ are coordinates on $Q$.  The Lagrangian density for the minimal immersions is the volume form on $X$ associated with the metric $h=\eta^*g$:
\[
\mathcal{L}:J^1P\rightarrow\Lambda^{n+1}X,\quad \cL\lp j^1\sig\rp=\sqrt{|h|}d^{n+1}x =: L( j ^1 \sigma)d^{n+1}x ,
\]
where $|h| := |\det h|$. Locally, the Lagrangian reads
\[
L\lp x_i,q_\alpha,\nu_j^\beta\rp = \left| \det \lp g_{\al \be}(q)\nu^\al_i\nu^\be_{j}\rp\right|^{\half},
\]
where $(x_i,q_\alpha,\nu_j^\beta) $ are the natural coordinates on $J ^1 P$.
\par
Since the bundle $P$ is trivial, the covariant Euler-Lagrange equations read
\begin{equation}\label{cov_E_L_trivial}
\frac{\delta L}{\delta \eta}-\operatorname{div}^Q\frac{\delta L}{\delta j^1\eta}=0,
\end{equation}
where $\operatorname{div}^Q$ is the divergence operator associated to a $\nabla^Q$-derivative (see \eqref{nabla_tilde_general_def}). Since $Q$ is a pseudo-Riemannian manifold, the Levi-Civita connection provides a natural choice of covariant derivative on $TQ$. The covariant Euler-Lagrange equations may be calculated by use of the following formula for the derivative of the determinant of an invertible matrix
\begin{equation}\label{der_det}
D \det (K) \cdot \delta K = \lp \det K\rp  \tr \lp K^{-1} \delta K\rp.
\end{equation}
Since $\nabla^Q$ is the Levi-Civita covariant derivative, the first term of \eqref{cov_E_L_trivial} vanishes and the covariant Euler-Lagrange equations have the local representation
\begin{equation}
\tilde \nabla_i^Q \;p^i_\al = 0,
\label{eq:ELMinImm}
\end{equation}
where
\[
p^i_{\al} = \dede{L}{\nu^\al_i} = \operatorname{sign}(h) |h|^\half h^{ij}\,g_{\al\be}\,\partial_j\eta^\be
\]
and $\tilde{\nabla}^Q $ is the $\nabla^Q$-derivative.

Note that when $X=\mR$ then equation \eqref{eq:ELMinImm} reduces to
\[
\tilde\nabla_t^Q \dot\eta = \nabla_{\dot\eta}^Q\dot \eta = 0
\]
which is just the geodesic equation on $(Q,g)$.

\medskip

Now consider the case when the isometry group $G=\operatorname{Iso}(Q,g)$ of $g$ acts freely and properly on $Q$.  Then $\pi_{M,Q}:Q\to M:=Q/G$ is a principal $G$-bundle.  Whereupon the group action $G\times P \to P$ by $\lp f, \lp x, q\rp\rp \mapsto \lp x, fq\rp$ gives $P$ a principal $G$-bundle structure over $\Sigma = X \times M$.  The geometric setup is as described in \S\ref{sec:geometric_setting}; this fact is elucidated by the following diagram:
\begin{diagram}
X\times Q &                       & \rTo^{\boldsymbol{p_1}} &                       & X \\
  & \rdTo_{\boldsymbol{\id \times \pi_{M,Q}}} &                  & \ruTo_{\boldsymbol{p_1}}  &   \\
  &                       & X\times M             &                       &   \\
\end{diagram}
The diagram also reveals that reduction of the minimal immersion problem is an example of {\em fiber bundle reduction}, the natural extension of the classical Lagrange-Poincar\'e reduction discussed in \S\ref{sec:fibre}.

Identification $\rho = \pi_{\Sig,P}\circ \sig \in \Gamma^{\operatorname{Imm}}\lp \pi_{X,\Sig}\rp$ with $r \in \operatorname{Imm}\lp X, M\rp$ may be effected by writing $\rho(x)=(x,r(x))$, where $r=\pi_{M,Q}\circ\eta$.  

\begin{lemma} The Lagrangian density $\mathcal{L}:J^1P\rightarrow \Lambda^{n+1}X$ defined by
\[
\mathcal{L}(j^1\sigma)= |h|^\half d^{n+1}x,
\]
is $G$-invariant.  Fixing a particular principal connection $\cA$ on $P$, the reduced Lagrangian $l:(J^1P)/G \to \mR$ may be expressed as
\[
l\lp j^1\rho, \bar\sig\rp = |h|^\half, \qquad h = r^*g_{M}\oplus \bar\sig^*g_{\Ad Q},
\]
where $r:=\pi_{M,Q}\circ\rho$, $g_M$ is the Riemannian metric on $M$ defined by $g_M = \lp\hor^{\cA}\rp^*g$, and $g_{\operatorname{Ad}Q}$ is the vector bundle metric on $\operatorname{Ad}Q$ defined by
\[
g_{\operatorname{Ad}Q}(m)\left(\llsb q,\xi\rrsb_{\mathfrak{g}},\llsb q,\xi\rrsb_{\mathfrak{g}}\right):=g(q)\left(\xi_Q(q),\eta_Q(q)\right).
\]
\end{lemma}
\begin{proof}  
Since $g$ is $G$ invariant in the sense that $f^*g = g$ for all $f \in G$,
\begin{align}
\cL\lp f\cdot j^1\sig\rp &= |\lp f\eta\rp^* g|^\half d^{n+1}x= |\eta^* f^* g|^\half d^{n+1}x\nonumber\\
&=|\eta^* g|^\half d^{n+1}x= \cL\lp j^1\sig\rp.
\label{eq:InvLagrangian}
\end{align}
Thus $\cL: J^1P \to \Lambda^{n+1}X$ is left $G$-invariant.

Since $Q$ is a pseudo-Riemannian manifold there exists a mechanical connections whose horizontal spaces are orthogonal complements of the vertical spaces.  This connection on $Q$ induces a unique connection $\mathcal{A}$ on $P$.  The reduced configuration space $J^1P/G$ is identified with $J^1\Sigma\oplus L\lp \pi^*_{X,\Sigma}TX,\Ad\,P\rp$ using the isomorphism $\beta_\mathcal{A}$ defined in \eqref{definition_beta}.  Consequently, the reduced Lagrangian $l:  J^1P/G\cong J^1\Sig \oplus_\Sig L\left(\pi_{X,\Sigma}^*TX,\operatorname{Ad}P\right) \to \mR$ may be expressed by
\begin{equation}\label{ReducedLagrangianMinImm}
l\lp \beta_{\cA}\lsb \gamma\rsb\rp d^{n+1}x = L\lp \gamma\rp,\quad \text{for all}\quad \gamma \in J^1P.
\end{equation}
Given a section $\sigma(x)=(x,\eta(x))$, the objective is to compute $h:=\eta^*g$ in terms of the reduced quantities $\rho(x)=(x,r(x))$ and $\bar\sigma(x)=\llsb\sigma(x),\sigma^*\mathcal{A}(x)\rrsb_\mathfrak{g}=\llsb\eta(x),\eta^*\mathcal{A}(x)\rrsb_\mathfrak{g}$, where in the last equality, the adjoint bundles of $P$ and $Q$ have been identified in the canonical way. Since $\mathcal{A}$ is the mechanical connection,
\[
g(q)(u_q,v_q)=g_M(\pi(q))(T\pi(u_q),T_q\pi(v_q))+g_{\operatorname{Ad}Q}(\pi(q))\left(\llsb q,\mathcal{A}(u_q)\rrsb_\mathfrak{g},\llsb q,\mathcal{A}(v_q)\rrsb_\mathfrak{g}\right).
\]
Thus,
\begin{align*}
\eta^*g(x)(u_x,v_x)&=g(\eta(x))(T_x\eta(u_x),T_x\eta(v_x))\\
&=g_M(r(x))(T_xr(u_x),T_xr(v_x))\\
& \qquad +g_{\operatorname{Ad}Q}(r(x))\left(\llsb \eta(x),\mathcal{A}\left(T_x\eta_x(u_x)\right)\rrsb_\mathfrak{g},\llsb \eta(x),\mathcal{A}\left(T_x\eta(v_x)\right)\rrsb_\mathfrak{g}\right)\\
&=\left(r^*g_M\right)(u_x,v_x)+g_{\operatorname{Ad}Q}(r(x))\left(\bar\sigma(u_x),\bar\sigma(v_x)\right),
\end{align*}
which is more compactly expressed by
\[
\eta^*g=r^*g_M\oplus \bar\sigma^*g_{\operatorname{Ad}Q}.
\]
This formula, together with \eqref{eq:InvLagrangian} and \eqref{ReducedLagrangianMinImm} combine to show that
\[
l(j^1\rho, \bar \sig) = |h|^\half, \qquad h = r^*g_M\oplus \bar\sig^*g_{\Ad}
\]
which is precisely the statement that was to be proved.
\end{proof}

\bigskip

Setting $\nabla^{\Sig}:=\nabla^{X}\oplus\nabla^M$, where $\nabla^M$ is the Levi-Civita connection associated to $g_M$ on $M=Q/G$ and $\nabla^X$ is an affine connection on $X$, formula \eqref{der_det} gives the functional derivatives,
\begin{align*}
\left\langle\dede{l}{j^1\rho},v\right\rangle &= \eval{\dd{}{\epsilon}}{\epsilon=0}l\lp j^1\rho + \epsilon v, \bar\sig\rp\\
&= \frac{1}{2}\operatorname{sign}(h)\sqrt{|h|}\,\tr\lp h^{-1}\eval{\dd{}{\epsilon}}{\epsilon=0} g_M(r)\lp Tr + \epsilon v, Tr + \epsilon v\rp\rp\\
&= \operatorname{sign}(h)\sqrt{|h|}\,\tr\lp \lp r^*g_M\rp^{-1}g_M(r)\lp Tr , v\rp\rp,
\end{align*}
and
\begin{align*}
\left\langle\dede{l}{\bar\sigma},w\right\rangle &= \eval{\dd{}{\epsilon}}{\epsilon=0}l\lp j^1\rho , \bar\sig+ \epsilon w\rp\\
&= \frac{1}{2}\operatorname{sign}(h)\sqrt{|h|}\,\tr\lp h^{-1}\eval{\dd{}{\epsilon}}{\epsilon=0} g_{\Ad Q}(r)\lp \bar\sig +\epsilon w, \bar\sig +\epsilon w\rp\rp\\
&=\operatorname{sign}(h)\sqrt{|h|}\,\tr\lp \lp\bar\sig^*g_{\Ad}\rp^{-1}g_{\Ad Q}(r)\lp \bar\sig, w\rp\rp.
\end{align*}
Derivation of the final functional derivative, $\de l/\de \rho$, requires a curve $\sig\lp\epsilon\rp \in \Sig$.  The horizontal curve 
\[
\sig^h_{\lp j^1\rho(x), \bar\sig(x)\rp}\lp\epsilon\rp \in J^1\Sig \oplus L\lp \pi_{X,\Sig}^*TX, \Ad\,P\rp
\]
then denotes the horizontal lift of $\sig(\epsilon)$ with respect to the affine connection $\nabla^{J^1\Sig}\oplus \nabla^{\cA}$. The final functional derivative is then defined by the relation
\[
\left\langle\dede{l}{\rho},u\right\rangle\lp x\rp =\eval{\dd{}{\epsilon}}{\epsilon=0}l\lp \sig^h_{\lp j^1\rho(x), \bar\sig(x)\rp}\lp \epsilon\rp\rp,
\]
as in \eqref{hor_deriv}.  Since $\nabla^M$ is the Levi-Civita connection with respect to $g_M$ there is no contribution to $\de l/ \de \rho$ from the $g_M$ terms of $l$.  Thus,
\begin{align*}
\left\langle\dede{l}{\rho},u\right\rangle\lp x\rp &=\eval{\dd{}{\epsilon}}{\epsilon=0}l\lp \sig^h_{\lp j^1\rho(x), \bar\sig(x)\rp}\lp \epsilon\rp\rp,\\
&= \frac{1}{2}\operatorname{sign}(h)\sqrt{|h|}\,\eval{\dd{}{\epsilon}}{\epsilon=0} \tilde\cR
\end{align*}
where
\[
\tilde\cR = \tr\lp \lsb \lp \bar\sig^h_{\lp j^1\rho(x), \bar\sig(x)\rp}\rp^*g_{\Ad Q}(r) \lp 0\rp \rsb^{-1} \lsb\lp \bar\sig^h_{\lp j^1\rho(x), \bar\sig(x)\rp}\rp^*g_{\Ad Q}(r_\epsilon) \lp \epsilon\rp \rsb\rp.
\]
Note that the first factor in the trace is evaluated at $\epsilon = 0$ {\em before} the derivative is taken.  Since $\tilde\nabla_\epsilon^\cA \bar\sig^h_{\lp j^1\rho(x),\bar\sig(x)\rp} = 0$ by definition, $\tilde\cR \in \sig^*\cF\lp \Sig\rp$.  Therefore introducing $\cR \in \cF\lp \Sig\rp$ such that $\sig^*\cR = \tilde\cR$ gives
\begin{align}
\eval{\dd{}{\epsilon}}{\epsilon = 0} \tilde\cR = u\contract d^M \cR.
\end{align}
This leads to the formula
\[
\scp{\dede{l}{\rho}}{u}= \frac{1}{2}\operatorname{sign}(h)\sqrt{|h|}\, u\contract d^M \cR.
\]
All the functional derivatives collected together read
\begin{equation}
\left\{
\begin{array}{rcl}\displaystyle
        \dede{l}{\bar\sig} &=& \operatorname{sign}(h)\sqrt{|h|} \lp\bar\sig^*g_{\Ad Q}\rp^{-1}\lsb \sig, \sig^*\lp\mI\cA\rp\rsb_{G}=:\bar\Pi\\
        \\
        \displaystyle\dede{l}{j^1\rho} &=& \operatorname{sign}(h)\sqrt{|h|} \lp r^*g_{M}\rp^{-1}g_{M}\lp r\rp\lp Tr,\cdot\rp =:P\\
        \\
        \displaystyle\dede{l}{\rho} &=& \frac{1}{2}\operatorname{sign}(h)\sqrt{|h|}\, d^M \cR,
\end{array}
\right.
\label{eq:RedMinImmDerivatives}
\end{equation}
where $\mathbb{I}:Q\times \mathfrak{g}\rightarrow\mathfrak{g}^*$ is defined by
\[
\langle\mathbb{I}(q)\xi,\eta\rangle:=g(\xi_Q(q),\eta_Q(q)) = g_{\Ad \,P}\lp \llsb q, \xi\rrsb_{\mg}, \llsb q, \eta\rrsb_{\mg}\rp.
\]
Now, restricting attention to a local neighborhood $U \subset X$  consider $\sig$ as a local section of the principal bundle $P^\rho$ over $U$.  Using capital letters for $\mg$ coordinates which are raised and lowered by $\mI$, lower case letters for $X$ coordinates raised and lowered by $h$ and greek letters for $M$ coordinates raised and lowered by $g_{M}$ results in the following local representations:
\begin{equation}
\left\{
\begin{array}{l}
\vspace{0.2cm}\displaystyle\lp\dede{l}{\bar\sig}\rp^i_J =C\lsb \sig, \lp\sig^*\cA\rp^i_{J}\rsb_G \cong C\lp\sig^*\cA\rp_{J}^i=:\Pi^i_J\\
\vspace{0.2cm}\displaystyle\lp\dede{l}{j^1\rho}\rp^i_\al = Cr^{,i}_\al=:P^i_\al\\
\displaystyle\lp\dede{l}{\rho}\rp_\al = \frac{1}{2C}\Pi^{Ki}\Pi_{K}^j\lp\Pi^{J}_{i}\Pi_{jJ}\rp_{,\al},
\end{array}
\right.
\label{eq:LocRedMinImmDerivatives}
\end{equation}
where
\[
C:= \operatorname{sign}(h)\sqrt{|h|}.
\]
The Lagrange-Poincar\'e equations read
\begin{equation}
\left\{
\begin{array}{l}\vspace{0.2cm}\displaystyle
\dive^MP = \frac{C}{2}\tr\lp \lp \bar\sig^*g_{\Ad Q}\rp^{-1}d^M\lp \bar\sig^*g_{\Ad Q}\rp\rp - \scp{\bar\Pi}{{\bf i}_{Tr}\tilde\cB}\\
\displaystyle\dive^\cA\bar\Pi - \ad^*_{\bar\sig}\bar\Pi = 0,
\end{array}
\right.
\label{eq:LPMinImm}
\end{equation}
where $\tilde\cB$ is the reduced curvature form associated to the connection $\cA$.  Equations \eqref{eq:LPMinImm} can be written in the local coordinates as
\begin{equation}\left\{
\begin{array}{l}
\displaystyle\vspace{0.2cm}\tilde\nabla^M_i P^i_\al =  \frac{1}{2C} \lp\Pi^{J\gam}\Pi_{J\gam}\rp_{,\al} + \Pi^j_JB^J_{\al \be}r^\be_{,j}\\
\displaystyle\pp{\Pi^i_{J}}{x^i} =0,
\end{array}
\right.
\label{eq:LocLPMinImm}
\end{equation}
where $B^J_{\al\be}$ is the local representation of the curvature form $\cB$ associated to $\cA$.  The right hand side of the first of equations \eqref{eq:LocLPMinImm} measures the deviation of $r$ being a minimal immersion in $M=Q/G$ whilst the second of equations \eqref{eq:LocLPMinImm} is Noether's Theorem.

For reconstruction, consider the form $\om^{\bar\sig}$ defined by equation \eqref{VerticalReconstructionForm}.  The curvature relation \eqref{reconstruction} is given in coordinate-free form as
\[
\mathcal{B}-\mathbf{d}^\mathcal{A}\omega^{\bar\sigma}-\frac{1}{2}\omega^{\bar\sigma}\wedge\omega^{\bar\sigma}=0,
\]
which is expressed in local coordinates by
\begin{equation}
B^J_{\al\be}= \lp\om^{\bar\sig}\rp^J_{\al,\be} - \frac{1}{C}c^J_{KM}\Pi^K_\al\lp\om^{\bar\sig}\rp^M_\be - \lp\om^{\bar\sig}\rp^J_{\be,\al} + \frac{1}{C}c^J_{KM}\Pi^K_\be\lp\om^{\bar\sig}\rp^M_\al + c^J_{KM}\lp\om^{\bar\sig}\rp^K_\al\lp\om^{\bar\sig}\rp^M_\be
\label{eq:LocMinImmCurv}
\end{equation}
where $c^J_{KM}$ are the structure constants for $\mg$.  The Reconstruction Theorem \ref{ReconstructionTheorem} states that if one can solve equations \eqref{eq:LocLPMinImm} and \eqref{eq:LocMinImmCurv} then there exists a unique solution of \eqref{eq:ExplicitReconstruction} which is also a solution of \eqref{eq:ELMinImm}.

The degree of geometric content in equations \eqref{eq:LPMinImm}-\eqref{eq:LocMinImmCurv} indicates why the problem of minimal immersions has fascinated mathematicians and other scientists for so long. One may expect that further investigations of this problem will continue to produce rich mathematical results.

\subsection{Metamorphosis image dynamics}\label{Met_im_dyn}

The metamorphosis framework is an interesting approach to the control theory problem of how best to match one image to another, particularly when the image possess attributes such as color, or some other representation of physical data.  This problem has applications in medical imaging where clinicians seek the best available tools to perform surgery in a non-invasive manner.  The metamorphosis approach to this problem was formulated in \cite{HoTrYo2008}.  This approach can be cast into the bundle picture as follows.
\par
Let $\mathcal{N}$ be a manifold of deformable objects (i.e. possible images) on a manifold $Q$.  For example, $\mathcal{N} = \operatorname{Emb}\lp M, Q\rp$, the embeddings of a manifold $M$, into $Q$, or $\mathcal{N} = \operatorname{Imm}\lp M, Q\rp$, the immersions of $M$ into $Q$.  Suppose that the diffeomorphism group $\mathcal{G}:=\operatorname{Diff}(Q)$ of $Q$ acts on $\mathcal{N}$ and consider the trivial fiber bundle $P = \mR \times \lp \mathcal{N} \times \mathcal{G}\rp\rightarrow\mathbb{R}$ on which the diffeomorphism group acts on the \textit{right\/} by the action
\[
(t, \eta, g)h :=(t,h^{-1}\circ \eta,h\circ\eta).
\]
The projection is given by
\[
\pi_{\Sig,P}:\mR \times \lp \mathcal{N} \times \mathcal{G}\rp\rightarrow\Sigma:=\mathbb{R}\times\mathcal{N},\quad \pi_{\Sig,P}(t, \eta, g)= (t,g\circ \eta ) = :(t, n).
\]
In the context of metamorphosis, $\eta$ is called the \textit{template}, $g$ the \textit{deformation} and $n=g\circ \eta$ the \textit{image}. Note that $X = \mR$ and this is an example of classical Lagrangian reduction (see particular case {\bf iv} above), for a given Lagrangian $L:\mathbb{R}\times T\mathcal{N}\times T\mathcal{G}\rightarrow\mathbb{R}$. In the context of metamorphosis, the Lagrangian does not depend on time and is given by
\begin{equation}\label{Lagrangian_metamorph}
L\lp g,\dot g, \eta, \dot \eta\rp  = \frac{1}{2}\| \dot g\|_\mathcal{G}^2 + \frac{1}{2\tau^2}\|Tg\circ \dot \eta\|_{\cN}^2
\end{equation}
where $\tau \in \mR$ is a parameter, $\|\!\cdot\!\|_\mathcal{G}$ denotes a $\mathcal{G}$-invariant metric on $T\mathcal{G}$, and $\|\!\cdot\!\|_\mathcal{N}$ denotes a metric on $T\mathcal{N}$.

The Lagrangian in this context is interpreted as the cost of using the controls and the aim is to minimize
\[
S = \int_0^1 L(\eta,\dot\eta,g,\dot g)\; dt,
\]
where the initial image $n_0=g_0\circ\eta_0$ and the final image $n_1=g_1\circ\eta_1$ are given.

The convective velocity
\begin{equation}\label{Flat-connection}
\cA\lp v_t,v_\eta,v_g\rp = g^{-1}v_g
\end{equation}
provides a suitable connection form.  Applying (\ref{beta_A_special}) yields
\begin{align*}
\beta_{\mathcal{A}}\lsb (t, \eta_t, g_t, \dot \eta_t, \dot g_t) \rsb_{\mathcal{G}} &= \left(T\pi_{X,P}(t, \eta_t, g_t, \dot \eta_t, \dot g_t),\llsb (t, \eta_t,g_t), \cA(t, \eta_t, g_t, \dot \eta_t, \dot g_t) \rrsb_{\mathfrak{g}}\right)\\
&= (t, n_t, \dot n_t, u_t)
\label{eq:imageiso}
\end{align*}
where $u_t = \dot g_t\circ g_t^{-1}$, $n_t=g_t\circ \eta_t$, and $\mathfrak{g}=\mathfrak{X}(Q)$ is the Lie algebra of the diffeomorphism group $\mathcal{G}$. Note that the time derivative of $n_t$ is given by the formula $\dot n_t=u_t\circ n_t+Tg_t\circ \dot\eta_t$. Denoting the reduced Lagrangian associated to $L$ by $l=l(t,n,\dot n,u):\mathbb{R}\times T\mathcal{N}\times\mathfrak{g}\rightarrow\mathbb{R}$ the Lagrange-Poincar\'e equations in this case read
\begin{equation}\label{eq:imageeq}
\left\{
\begin{array}{l}
\vspace{0.2cm}\displaystyle\partial_t \dede{l}{u_t} + \pounds_{u_t} \dede{l}{u_t} = 0
,\\
\displaystyle\frac{D^\nabla}{Dt} \dede{l}{\dot n_t} - \dede{l}{n_t} = 0,
\end{array}
\right.
\end{equation}
where $\nabla$ is a torsion free covariant derivative on $\mathcal{N}$ and the fact that the group action is a {\it right} action is carefully noted. The \emph{reduced}  Lagrangian associated with $L$ in \eqref{Lagrangian_metamorph} reads
\[
l(n,\dot n,u)=\frac{1}{2}\|u\|^2_\mathcal{G}+\frac{1}{2\tau^2}\|\dot n-u\circ n\|^2.
\]
Equations \eqref{eq:imageeq} are equivalent to those in \cite{HoTrYo2008}; but they are simpler in form and  expressed in different variables. In these variables, equations \eqref{eq:imageeq} have split into horizontal and vertical parts with respect to the flat connection (\ref{Flat-connection}), thereby resulting in their zero right-hand sides.

\subsection{Covariant formulation of  metamorphosis image dynamics}\label{CovMeta}

Metamorphosis image dynamics for immersions may be placed into a covariant setting. This is achieved by replacing $X=\mR$ by $X= \mR \times M$ and $\operatorname{Imm}(M,Q)$ by $Q$, and by considering the trivial fiber bundle
\[
\pi_{X,P}:P= X \times \lp Q \times \mathcal{G}\rp\rightarrow X,\quad (t,m,q,g)\mapsto (t,m)=:x.
\]
Now, let the diffeomorphism group $\mathcal{G}=\operatorname{Diff}(Q)$ act on $P$ by the \textit{right} action
\[
(x, q, g)h = (x, h^{-1}(q), g\circ h),
\]
and obtain the principal bundle
\[
\pi_{\Sig,P}:P= X \times \lp Q \times \mathcal{G}\rp \rightarrow\Sigma=X\times Q,\quad \lp x, q, g \rp \mapsto (x, g(q))=:(x,n).
\]
In this framework,  templates and deformations are sections of $\pi_{X,P}$ where the deformation is further specified to be independent of the variable $m\in M$.  Concretely, for $\sig \in \Gamma\lp \pi_{X,P}\rp$,
\begin{equation*}
\sig\lp t,m\rp = \lp t,m, \eta\lp t,m\rp, g_{(t,m)}\rp,
\end{equation*}
where $\eta:X\rightarrow Q$ and $g:X\rightarrow\mathcal{G}=\operatorname{Diff}(Q)$.

The restrictions required to mimic the classical metamorphosis are:
\begin{equation}\label{condition_imm}
\eta_t:=\eta\lp t, \cdot\rp \in \operatorname{Imm}(M,Q)\quad\text{and}\quad  g_{(t,m_1)} = g_{(t,m_2)}, \quad\text{for all}\quad  m_i \in M.
\end{equation}
For the first restriction no constraint is necessary since our the requirement is that
\begin{equation}
\operatorname{rank} \left(T^M_m\eta_t\right) = \dim M,\quad\text{for all}\quad t\in\mathbb{R},
\end{equation}
where $T^M\eta_t:TM\rightarrow TQ$ denotes the tangent map of $\eta_t$, the variable $t$ being considered as a parameter.

The first condition in \eqref{condition_imm} formally defines an open subset of the space of curves with values in the manifold $\mathcal{F}(M,Q)$ of smooth maps from $M$ into $Q$. To see this, recall that the space $\operatorname{Imm}(M,Q)$ of immersions is an open subset of $\mathcal{F}(M,Q)$ and that 
$\Gamma\lp \pi_{X,\Sigma}\rp$ may be identified with the space of curves in $\mathcal{F}(M,Q)$.  

The second condition in \eqref{condition_imm} is enforced by a Lagrange multiplier. Consequently, the new Lagrangian reads:
\[
\tilde L\lp j^1\sig, \lambda\rp = L\lp j^1\sig\rp - \dscp{\lambda}{(T^Mg)\circ g^{-1}},
\]
where the Lagrange multiplier is a section
\[
\lambda \in  \Gamma\lp \pi_{X,L(T^*M,\mathfrak{g}^*)}\rp,\quad\text{where}\quad\mathfrak{g}^*=\Omega^1(Q),
\]
and $\dscp{\cdot}{\cdot}$ denotes the pairing between the spaces $L(T^*M,\mathfrak{g}^*)$ and $L(TM,\mathfrak{g})$.
Note that here $T^Mg$ denotes differentiation of the section dependence on $M$, {\em not} the argument of the diffeomorphism in $Q$.

One creates analogous Lagrangians as before, where now the spatial derivatives of $\eta$ are considered to be independent variables in the theory.  That is,
\[
j^1\sig = \lp x, \eta, g, \dot\eta\,dt + T^M\eta, \dot g\,dt+T^Mg\rp.
\]
An example of a possible Lagrangian for metamorphosis is
\[
L\lp g, \dot g, T^Mg, \eta, \dot \eta, T^M\eta\rp = \frac{1}{2}\|\dot g\|^2_\mathcal{G} + \frac{1}{2\tau^2}\|Tg(\dot \eta) \|^2_Q + \frac{1}{2\kappa^2} \|Tg\!\cdot\!T^M\eta\|^2_L\,,
\]
where $\|\!\cdot\!\|_Q$ is norm associated to a metric on $Q$ and $\|\!\cdot\!\|_Q$ is associated to a vector bundle metric on $L(TM,TQ)$. Note that in the previous formula the Lagrangian is interpreted as being defined on an arbitrary element of the first jet bundle $J^1P$ and not necessarily on the first jet extension of a section. Therefore, $\dot\eta$ denotes an arbitrary element in $T_\eta Q$ and $T^M\eta$ is an arbitrary element in $L(T_mM,T_\eta Q)$.

A diffeomorphism $h\in\mathcal{G}$ acts on the first jet extension $j^1\sigma$, as follows:
\[
\lp x, \eta, g, \dot\eta\,dt + T^M\eta, \dot g\,dt+T^Mg\rp\mapsto \lp x,h^{-1}\circ \eta,g\circ h, Th^{-1}\circ(\dot \eta\, dt+T^M\eta),(\dot g\, dt+T^Mg)\circ h\rp;
\]
thus the Lagrangian $L$ is $\mathcal{G}$-invariant.

Fixing the connection $\cA\lp v_x, v_q, v_g\rp =g^{-1} v_g$ and writing $(t,m)=x$ yields
\begin{align*}
\bar\sig (x)&= \llsb \sig(x), \sig^*\cA\rrsb_{\mathfrak{X}(Q)} =\left [ \!\! \left[ (x,\eta(x),g_x),g_x^{-1}(\dot g_x+T^Mg_x)   \right ] \!\! \right ]_{\mathfrak{X}(Q)}\\
& = \lp x,n(x), \dot g_xg_x^{-1}  dt + \lp T^Mg_x\rp \circ g_x^{-1}\rp,
\end{align*}
where $n(x) = g_x(\eta(x))$ which is the same definition of $n$ as in \S\ref{Met_im_dyn}.  The Lagrange-Poincar\'e equations can now be written in the convective representation from \S\ref{sec:Representations} as
\[
\left\{
\begin{array}{l}
\displaystyle\vspace{0.2cm}\prt_t \dede{l}{u_t} + \dive^M\lp \dede{l}{\lp T^Mg\rp\circ g^{-1}} - \lambda\rp + \pounds_{u_t} \dede{l}{u_t} =0,\\
\displaystyle\vspace{0.2cm}\dive^Q \left(\dede{l}{ j^1n}\right) - \dede{l}{n} =0,\\
\displaystyle\lp T^Mg \rp \circ g^{-1}=0.
\end{array}
\right.
\]
Seeking a solution with 
\[
\lambda = \dede{l}{\lp T^Mg\rp \circ g^{-1}}
\]
yields the equations
\begin{equation}\label{eq:covimageeq1b}
\left\{
\begin{array}{l}
\displaystyle\vspace{0.2cm}\prt_t \dede{l}{u_t} + \pounds_{u_t} \dede{l}{u_t}= 0,\\
\displaystyle\dive^Q 
\left(\dede{l}{ j^1n}\right) - \dede{l}{n}= 0.
\end{array}
\right.
\end{equation}
Note that the first equation above is identical to the first equation in \eqref{eq:imageeq} whilst the second equation \eqref{eq:covimageeq1b} is a covariant analogue of the second equation in \eqref{eq:imageeq}.

This covariant formulation takes the problem from classical Lagrangian reduction (as in particular case {\bf iv}), to a covariant problem that does not fall into any of the particular examples.

One of the potential advantages of such a transformation would be to apply multisymplectic integrators.  The infinite-dimensional approach introduces special dependence on the time variable and is formulated on an infinite-dimensional manifold.  The covariant formulation, on the other hand, takes advantage of exchange symmetry $t\leftrightarrow m_j$ for any component $m_j$ of $m$ to formulate the problem in a multisymplectic fashion.  If there is no coupling in the Lagrangian between the fibers of $\Ad P$ and $J^1\Sig$, then the multisymplectic problem is posed on a finite dimensional manifold.

\section{Conclusion and future directions}\label{Conclusion}

This paper has presented a framework for Lagrange-Poincar\'e reduction that unifies the approaches taken in the particular cases  {\bf i} - {\bf iv} and extends the Lagrange-Poincar\'e theory beyond the scope of those cases.  On one hand, the work of \cite{CaRaSh2000} and \cite{CaRa2003} has been extended to apply to the general fiber bundle case.  On the other hand, the classical Lagrange-Poincar\'e theory developed in \cite{CeHoMaRa1998} and \cite{CeMaRa2001} has been extended to the field theoretic setting.  

Two surprising results have appeared that concern the integrability conditions associated with both reconstruction and the convective representation of the dynamics.  First, the requirement of an additional condition for reconstruction first observed in \cite{CaRa2003} was found also to occur here, even though less geometric structure is present.  Second, the convective representation may not always exist for an  arbitrary problem. This observation highlights the importance of the geometric tools used in formulating the framework. Even though the convective representation may not exist, the general $\Ad P$-valued objects do exist and they can be used either to study the dynamics or to find an alternative representation.

Also note the large range in applications: the examples given here range from the \textit{classical Skyrme model} in physics to \textit{molecular strand dynamics} in biology, from \textit{metamorphosis image dynamics} in computer science and control theory to the \textit{isoperimetric problem} in mathematics.  This range of examples is compelling and an accurate reflection of the unifying power of the framework developed.

The covariant metamorphosis example in \S\ref{CovMeta} has revealed an interesting property of the Lagrange-Poincar\'e theory. Namely, the covariant expression of existing classical Lagrange-Poincar\'e problems in the present framework produces a geometric reformulation of the problem.  Further investigation into this process of geometric reformulation could lead, for example, to new applications of multisymplectic integrators.

Finally, the Kelvin-Noether Theorem has been extended to the Lagrange-Poincar\'e field setting.  This extension is two-fold.  Firstly, the Kelvin-Noether theorem is usually stated for Euler-Poincar\'e systems where there is no shape space, the Theorem now to applies to Lagrange-Poincar\'e systems where a shape space is present.  Secondly, the Kelvin-Noether Theorem now extends from the classical context to the covariant context.  This result is particularly important, since it constitutes a major tool for gaining qualitative information about any problem formulated within the scope of the Lagrange-Poincar\'e field framework.

{\footnotesize

\bibliographystyle{newapa}
\addcontentsline{toc}{section}{References}
\bibliography{references}
\rem{

}
}

\end{document}